\documentclass[review,3p,times,10pt,authoryear]{elsarticle}
\usepackage{amssymb}
 \usepackage{lineno}
 \usepackage{float}
 \usepackage[english]{babel}
 \usepackage[utf8]{inputenc}
\usepackage{enumitem} 
\usepackage{makecell}
 \usepackage{comment}
\usepackage[hyphens]{url}
\usepackage{hyperref}
\hypersetup{colorlinks=true,breaklinks=true}
\usepackage{graphicx}
\usepackage{longtable}
\usepackage{multirow}
\usepackage{caption}
\usepackage{subcaption}

\usepackage{pgf-pie}

\usepackage{pgfplots}
\pgfplotsset{width=20cm,compat=1.8}
\usetikzlibrary{shadows}

\usepackage{lineno}

\journal{Intelligent Systems with Applications}

\begin{document}

\begin{frontmatter}
\title{Smartphone Apps for Tracking Food Consumption and Recommendations: Evaluating Artificial Intelligence-based Functionalities, Features and Quality of Current Apps\footnote{Cite this paper as: Samad, S., Ahmed, F., Naher, S., Kabir, M. A., Das, A., Amin, S., \& Islam, S. M. S. (2022). Smartphone Apps for Tracking Food Consumption and Recommendations: Evaluating Artificial Intelligence-based Functionalities, Features and Quality of Current Apps. Intelligent Systems with Applications, 200103. \url{https://doi.org/10.1016/j.iswa.2022.200103}}}

\author[inst1]{Sabiha Samad}\ead{u1604035@student.cuet.ac.bd}
\author[inst1]{Fahmida Ahmed}\ead{u1604107@student.cuet.ac.bd}
\author[inst1]{Samsun Naher}\ead{u1604048@student.cuet.ac.bd}
\author[inst2,inst3]{Muhammad Ashad Kabir\corref{c1}}
\cortext[c1]{Corresponding author
}
\ead{akabir@csu.edu.au}
\author[inst4]{Anik Das}\ead{x2021gmg@stfx.ca}
\author[inst4]{Sumaiya Amin}\ead{x2020gae@stfx.ca}
\author[inst5]{Sheikh Mohammed Shariful Islam}\ead{shariful.islam@deakin.edu.au}
\affiliation[inst1]{organization={Department of Computer Science and Engineering, Chittagong University of Engineering and Technology},
            city={Chattogram},
            postcode={4349}, 
            country={Bangladesh}}
\affiliation[inst2]{organization={School of Computing, Mathematics and Engineering, Charles Sturt University},
            city={Bathurst},
              state={NSW},
            postcode={2795},
            country={Australia}}

\affiliation[inst3]{organization={Gulbali Institute for Agriculture, Water and Environment, Charles Sturt University}, city={Wagga Wagga}, state={NSW}, postcode={2678},  country={Australia}
}

\affiliation[inst4]{organization={Department of Computer Science},
            addressline={St. Francis Xavier University}, 
            city={Antigonish},
            postcode={B2G 2W5}, 
            state={NS},
            country={Canada}
}
\affiliation[inst5]{organization={Institute for Physical Activity and Nutrition, Deakin University}, 
city={Geelong}, state={VIC}, postcode={3216}, country={Australia}
}
             
\begin{abstract}
The advancement of artificial intelligence (AI) and the significant growth in the use of food consumption tracking and recommendation-related apps in the app stores have created a need for an evaluation system, as minimal information is available about the evidence-based quality and technological advancement of these apps. Electronic searches were conducted across three major app stores and the selected apps were evaluated by three independent raters. A total of 473 apps were found and 80 of them were selected for review based on inclusion and exclusion criteria. An app rating tool is devised to evaluate the selected apps. Our rating tool assesses the apps’ essential features, AI based advanced functionalities and software quality characteristics required for food consumption tracking and recommendations, as well as their usefulness to general users. The rating tool's internal consistency, as well as inter- and intra-rater reliability among raters, are also calculated. Users' comments from the app stores are collected and evaluated to better understand their expectations and perspectives. Following an evaluation of the assessed applications, design considerations that emphasize automation-based approaches using artificial intelligence are proposed. According to our assessment, most mobile apps in the app stores do not satisfy the overall requirements for tracking food consumption and recommendations. ``Foodvisor" is the only app that can automatically recognise food items, and compute the recommended volume and nutritional information of that food item. However, these features need to be improvised in the food consumption tracking and recommendation apps. This study provides both researchers and developers with an insight into current state-of-the-art apps and design guidelines with necessary information on essential features and software quality characteristics for designing and developing a better app.
\end{abstract}
\end{frontmatter}

\section{Introduction}\label{intro}


Food is one of the most basic requirements of human life. It is often regarded as much more than a means of survival, and proper food intake is essential for human health and fitness. Our health is closely dependent on the 4 types or amount of food we intake~\citep{min2019survey}. There are numerous fields such as sociology, psychology, nutrition sciences, and medicine in which healthy food consumption is explored~\citep{mai2017indirect}. Food choices are negatively influenced by a busy lifestyle, bad habits, and low self-control~\citep{brug1995psychosocial,koenigstorfer2014healthful}. However, excessively unhealthy lifestyles and bad dietary habits, such as increased food intake with high energy and high fat, lead to various health issues~\citep{ng2014global}. According to the World Health Organization (WHO), more than 1.9 billion adults (aged over 18) are overweight, and more than 650 million people suffer from obesity~\citep{chu2018update}. Many chronic diseases such as hypertension, type 2 diabetes mellitus, cardiovascular disease, and stroke are linked to obesity and excess weight~\citep{speiser2005obesity}. This problem is becoming a significant health concern.  
One of the main reasons for the obesity problem is that many people follow a very unhealthy lifestyle. Their dietary habits are also unhealthy, such as increased food intake with high energy and high fat. The intake of highly caloric, inexpensive, larger portion sizes and nutrient-dense foods promoted by environmental changes, coupled with decreased physical activity, and increased sedentary behaviors, is a significant causative factor for obesity~\citep{beal2013should}.

 In recent years, the use of smartphones to track food consumption or compute the nutritional value of food's has expanded due to the increasing number  of food consumption tracking and recommendation apps in the app stores, and the great potential of smartphone's to be a useful tool~\citep{KALINOWSKA2021271}. Nowadays in app stores, many apps are focused on health and fitness. In the major app stores, there were 32500 mobile health apps available in 2017 and this number is continuing to rise~\citep{ferrara2019focused}. Apps can play an important role in simplifying the tracking of health-related behaviors and weight management~\citep{chen2015most}. Moreover, the usage of smartphones and rapid development of artificial intelligence (AI) technologies have enabled new food identification systems for dietary assessment, which are significant for the prevention and treatment of chronic diseases such as type 2 diabetes mellitus, cardiovascular disease, and overcoming health issues such as obesity~\citep{min2019survey}. Furthermore, food intake behaviour (e.g., assessment of calorie intake, nutritional analysis, and eating habits) can be analyzed if food items or categories are recognized. 
 
 Recently, AI and machine learning based mobile food recognition methods are also being implemented. For example, \cite{he2014analysis} used AI techniques for identifying food from an image. The bag of visual words model (BoW) has been used for representing food images as visual words distributions and the support vector machine (SVM) model has been used to classify~\citep{farinella2014classifying}. Furthermore, \cite{anthimopoulos2014food} used SVM, artificial neural Network and random forest classifications on 5000 food images organized into 11 classes described in terms of different bag-of-features. The convolutional neural network (CNN) is also used in some studies~\citep{christodoulidis2015food,kawano2014food}. \cite{ming2018food} proposed a photo-based dietary tracking system that employed deep-based image recognition algorithms to recognize food and analyze nutrition. For estimating an individual's food and calorie intake, the calculation of food portion size or volume is necessary. In several studies, different types of methods (i.e., single image-based or multiple image-based) have been used for estimating food volume from food images~\citep{kong2012dietcam,sun2010wearable,dehais2016two,fang2018single,meyers2015im2calories}. To achieve quantitative food intake estimation, researchers combined visual recognition and 3D reconstruction in a study~\citep{puri2009recognition}. Both Android smartphone and web-based applications are implemented to recognize food and estimate the calorific and nutritional content of foods automatically without any user input~\citep{zhang2015snap}. 
 
 Food recommendation is a significant domain for people as well as society~\citep{min2019survey}. Incorporating health into recommendations is mostly a recent concern~\citep{rokicki2018impact,nag2017pocket,yang2017yum}. \cite{mokdara2018personalized} proposed integrating deep neural network with a recommendation system focusing on Thai food. It not only considers users' food choices but also pays attention to users' health. Based on individual customer behaviors, tastes, and eating history, the system will assist consumers in making food selection decisions. Besides, a food recommendation system has been built to recommend food to diabetic patients based on nutrition and food characteristics~\citep{phanich2010food}.


Reviews on various health-related apps have been conducted in many different studies. A prior study reviewed diet tracking apps common in the Apple App Store and Google Play Store~\citep{ferrara2019focused}. \cite{franco2016popular} analyzed the main features of the most popular nutrition apps and compared their strategies and technologies for dietary assessment and user feedback. Another study reviewed nutritional tracking mobile applications specifically for diabetes patients~\citep{darby2016review}. \cite{rivera2016mobile} characterized the use of evidence-based methods, the participation of health care experts, and the clinical assessment of commercial smartphone applications for weight loss or weight control. 
In this study, we evaluated the apps from three commercial app stores -- Google Play, Apple App Store, and the Microsoft Store -- to evaluate food consumption tracking and recommendation apps for all users, not just diabetes patients, pregnant women, or children.
To the best of our knowledge, no research has thoroughly examined the current commercial mobile app market landscape to analyze and scientifically evaluate apps linked to food consumption tracking and recommendations. The speedy growth of such apps in the app stores,- and the fast acceptance of these apps by the general population necessitates an assessment of this rapidly expanding market. 

In this study, we have conducted a critical review of food consumption tracking and recommendation apps accessible in the three major commercial app stores (i.e., Google Play Store, Apple App Store, and Microsoft Store). 
We found a total of 473 apps in our initial search; after excluding the apps based on our exclusion criteria, we finally selected 80 apps for our study. We devise an app rating tool by adopting and extending the existing app rating tools to assess those selected apps using three raters. The rating tool and the rating quality of raters are examined through internal consistency, and inter- and intra-rater reliability, respectively. We also analysed the user comments from app stores to better understand users' expectations and perspectives. We also discuss the limitations of the reviewed apps and potential design considerations from the perspectives of both developers and researchers. 


The rest of the paper is organized as follows. 
Section \ref{sec:methods} describes the methodology of our work, including the app search procedure, the measures used in app selection, and our devised app rating tool.
In Section \ref{result}, we present the results of our study that include the overall assessment of the apps, internal consistency of our rating tool, intra- and inter-rater reliability, analysis of app store ratings and our measured ratings, assessment of functionality criteria, and analysis of users' comments from app stores. In Section \ref{discus}, principal findings (including the limitations of the reviewed apps and design considerations) and the limitations of this study are discussed. Finally, Section \ref{conclusion} concludes the paper and outlines future research directions. 

\section{ Methods}\label{sec:methods}

\subsection{App search procedure} 
We have performed an electronic search to identify the relevant apps from three major commercial app stores, i.e., Google Play Store, Apple App Store, and Microsoft Store. Following similar approaches used in previous studies~\citep{rivera2016mobile,ashad2020mobile}, a keyword-based search process was used. The guidelines for Preferred Reporting Items for Systematic Reviews and Meta-Analysis (PRISMA)~\citep{tricco2018prisma} were followed to ensure transparency and clarity in reporting, as well as the ability for other researchers to replicate the search process. The keywords used in the search were specifically selected by studying the names of several prominent food consumption tracking and recommendation apps, so that the search would yield the same result if the same keywords were used at the same time and from the same location~\citep{stawarz2015beyond}. ``Food consumption", ``calorie consumption", ``daily food consumption", ``nutrition consumption", and ``track food consumption" are the keywords used for searching.

The investigators worked together to conduct the search, screening, and final inclusion process. All three app stores were searched using the same set of keywords to minimize variances and maintain uniformity. Three investigators independently conducted the same search using the same terms numerous times before compiling the final list of food consumption tracking and recommendation applications for inclusion. For each app store they searched, each investigator created their list of apps. They screened the apps using the inclusion and exclusion criteria (described in Section \ref{Measures used in apps’ selection}). This search and selection process was carried out by each investigator using their smartphone. The investigators' different lists were combined to create the final app list to be examined and analyzed for this study. Conflicts between the lists were resolved by a group discussion among all the investigators.




\subsection{Raters}
Expert raters were selected to rate all the apps. They included three final year Bachelor of Computer Science students with two years of mobile application development experience. Also, two computer science graduates with two years' of mobile application development experience rated three apps named ``Weight Loss Coach \& Calorie Counter - Nutright", ``Foodzilla! Nutrition Assistant, Food Diary, Recipe" and ``Fitatu Calorie Counter - Free Weight Loss Tracker" for measuring internal consistency.
All the apps on the investigators' final app list were rated separately by the raters. Their responses were collected in a response form (Google Forms), and the data from the spreadsheet attached to the form was used to extract the data from the raters.

\subsection{Measures used in apps’ selection}\label{Measures used in apps’ selection}
\begin{figure}[!htb]
\centering
\includegraphics[width=1\textwidth]{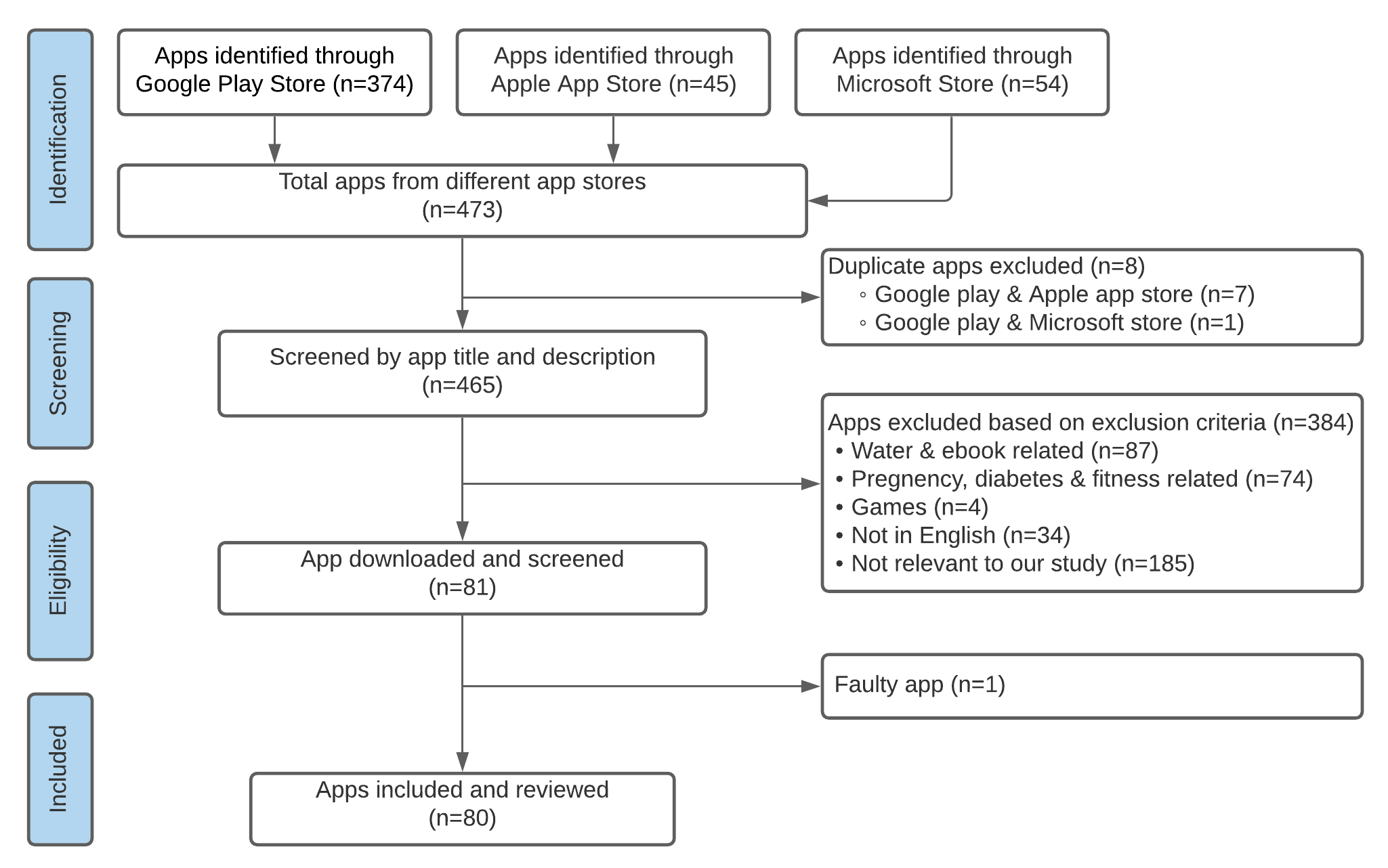}
\caption{Flow diagram of study methods}
\label{prisma}
\end{figure}

The methodology we used for the identification, screening, eligibility, and selection of apps is shown in Figure~\ref{prisma}.
During the search process, a keyword-based search technique was used on three app stores separately, yielding 473 apps.
Eight apps were removed because they had identical apps from the same developer or publisher in multiple app stores (a duplicated app). Before being eliminated, these apps were tested on their respective platforms (Android, iOS, and Windows) to see if they had the same functionality. The remaining 465 apps were screened based on their title and description.
After that, the apps were chosen based on their description in the app store. This was the first stage of the screening process. If the app description stated that it tracked food consumption or provided food recommendations, it was included in the study. For further curation, we considered the following inclusion criteria: (1) apps that can track food consumption, (2) apps that compute food portion size and estimate nutritional values, (3) apps that present consumption history visually, (4) apps that recommend food to the users. Also, we looked for apps that allow users to contribute new food items' names to their database. These criteria were applied to the selected apps to ensure that they met our study's requirements. If any of these features were contained in an app, we took those apps and included them in to our apps list. 
From the primary screening of the apps, apps were excluded for one or several of the following reasons: (1) apps that had functionality such as water consumption tracking and eBook related, (2) apps that were solely focused on pregnancy and baby food, diabetes, fitness and exercise, (3) food-related games, (4) apps in non-English languages, (5) apps that are not relevant to our study like fasting related, food photo sharing related, health tips related, and only recipe suggesting related, Sugar trackers, step counters, blood test guides, protein trackers, and wine consumption trackers were also excluded. In the secondary screening, all 81 remaining apps were downloaded and evaluated by each rater individually. One app was removed at this point because it was malfunctioning. In the end, 80 apps (70 from the Google Play store, 6 from the Microsoft Store and 4 from the Apple App store) selected for this study were analyzed and reviewed.

\subsection{App rating tool}
We have devised a rating tool for evaluating the selected apps and determining their appropriateness and usability.
We reviewed research on software quality aspects such as usability, reliability, functionality, and efficiency~\citep{koepp2020quality,poon2015algorithms,friesen2013mhealth,vos2013poster}. Our goal was to build a rating tool by adopting and extending the existing rating tools such as the mobile application rating scale (MARS)~\citep{stoyanov2015mobile}, uMARS -- end- user version of MARS~\citep{stoyanov2016mobile}, FinMARS -- MARS for financial apps~\citep{huebner2019finmars}, a mobile app rating tool for foot measurement~\citep{ashad2020mobile}, and a mobile app rating tool for child sexual abuse education~\citep{pritha2021systematic}. Our developed rating tool adopts the relevant evaluation features suited for analyzing food consumption tracking and recommendations. The finalized rating tool with the updated sub-scales and their respective criteria have been portrayed in ~\ref{sec:sample:appendix}. 

We have devised the app rating tool by clustering the domains according to app quality criteria. We used a Likert scale which is a popular instrument~\citep{wu2007empirical} ranging from 1 to 5 representing very bad to very good, respectively. For example, if an app can recognise food items from photos and also from an app databas, then we consider it as the highest quality feature and we rate the app as 5 for this feature. If an app can recognise a food item from a photo but not from a database, we rated the app as 4. We rated the app as 3 when it can recognise food from barcode scanning, and 2 when an app can recognise food from a database or allows manual entry. Lastly, if an app cannot recognise food by any means, we evaluated the app as 1. We applied this evaluation technique for every question of our food app rating tool. 

We also added a rating option labeled ``Unknown" because we were facing problems accessing certain information. For example, we could not find whether an app was open source or not, so we labeled those apps' sources as unknown. We used descriptive answers for app metadata items. These items are store name, app name, app rating, developer name, applicable age group items, and app sub-category items. But, we used the Likert scale for the rest of the questions. In the following subsections, we describe all the sub-scales of our app rating tool.

\subsubsection{App metadata}
Metadata is data that provides information about other data. App metadata has been clustered with the general information of the apps which were gathered from the respective app stores. Table~\ref{tab:lbl} reports metadata of our reviewed apps such as platform, country of origin, business model (free/paid), app rating, and number of downloads.
 
\begin{table}[!htb]
    \centering
    \caption{Apps metadata}
    \label{tab:lbl}
   {\small
    \begin{tabular}{lc}
    \hline
     Item  & \makecell[c]{Count (N=80)\\n (\%)} \\
    \hline\hline
    \textbf{Platform} & \\
    \quad Android & 70 (87.5\%)\\
    \quad Windows & 6 (7.5\%)\\
    \quad iOS & 4 (5\%)\\
  \textbf{Business model} & \\
  \quad Completely free &29 (36.25\%) \\
  \quad Limitedly free &51 (63.75\%) \\
    \textbf{Download} & \\
    \quad 50M+ & 1 (1.25\%)\\
    \quad 10M+ & 4 (5\%)\\
    \quad 5M+ & 2 (2.5\%)\\
    \quad 1M+ & 12 (15\%)\\
    \quad 500K+ & 7 (8.75\%)\\
    \quad 100K+ & 12 (15\%)\\
    \quad 50K+ & 8 (10\%)\\
    \quad 10K+ &8 (10\%)\\
    \quad 5K+ &2 (2.5\%) \\
    \quad 1K+ &7 (8.75\%) \\
    \quad 500+ &3 (3.75\%) \\
    \quad 100+ &3 (3.75\%)\\
    \quad Not available & 11 (13.75\%) \\
    \textbf{Country of origin} & \\
    \quad USA & 23 (28.75\%)\\
    \quad India & 8 (10\%)\\
    \quad UK & 5 (6.25\%) \\
    \quad France & 4 (5\%) \\
    \quad Ukraine, Russia, Germany & 3 each (3.75\%) \\
    \quad Canada, South Africa, Spain & 2 each (2.5\%) \\
    \quad \makecell[l]{Finland, Sweden, Poland, Serbia, Australia,\\\quad Denmark, Bulgaria, Singapore, Netherlands, \\\quad New Zealand, Switzerland, South Korea} & 1 each (1.25\%)\\
    \quad Unknown & 13 (16.25\%) \\
    
    \hline
    
    \hline
    \end{tabular}
    }
\end{table}

\subsubsection{App category}
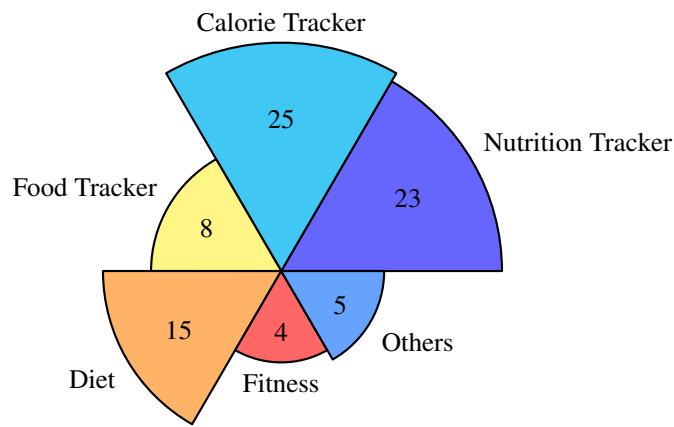
\begin{figure}[!htbp]
\centering
\resizebox{9cm}{!}{
\begin{tikzpicture}
  \pie[polar,sum=auto]{23/Nutrition Tracker, 25/Calorie Tracker, 8/Food Tracker, 15/Diet, 4/Fitness, 5/Others}
\end{tikzpicture}}
\caption{Categorical representation of the evaluated apps}
\label{category}
\end{figure}

All the included apps were divided into sub-categories like nutrition tracker, calorie tracker, food tracker, diet, fitness, and others, focusing on their main aim and functionalities (see Figure~\ref{category}). In the sub-category of nutrition tracker, the focus of the apps was tracking nutrition but some also tracked calorie consumption. The nutrition tracker sub-category contained 28.75\% (23/80) of the total apps. However, in the calorie tracker sub-category, 25 out of 80 apps (31.25\%) only tracked calorie consumption. Food tracker apps were focused on tracking food the names only. They merely tracked nutrition or calorie consumption. Only 10\% (8/80) of the total apps were listed in the food tracker sub-category. Our evaluation procedure, found some apps that can track nutrition or calorie consumption, but they mainly focused on suggesting a diet plan for users. We considered these apps in the diet sub-category; 18.75\% (15/80) of the apps were in this category(e.g., ``Keto Manager: Keto Diet Tracker \& Carb Counter" app). Also, we found 4 out of 80 apps focused on  improving their users' fitness (5\%) by tracking exercise, suggesting workout routines, and so on. However, those apps could also track calories or nutrition consumption (e.g., ``Fitstyle - Home Workout, Fitness \& Diet Plan"). We considered 5 out of the 80 apps (6.25\%) in the others category because most of them focused on multiple features, and some of these features matched our key features. For example, the app ``MealLogger-Photo Food Journal", is like social media for health-conscious people. Another example is ``Health Mate - Calorie Counter \& Weight Loss", which can track heart rate, sleep routine, shows food insights, and so on. Most apps were in the calorie tracker category, followed by. The nutrition tracker and diet plan focused app categories. The fitness focused,- and diet plan focused apps could also track calorie/nutrition/food, and show food consumption history. 

\subsubsection{Aesthetics}\label{sec:aesthetics}
Visual appeal is one of the key factors for the success of any app. This sub-scale criteria is equally important as the core functionalities and performance of an app~\citep{chetrari2017characteristics}. There are many apps in app stores that have similar functionalities, so visual appeal is often the main difference between them. In today’s competitive marketplace, whether an app will be successful or not, is widely dependant on the layout and organisation of the user interface components integrated into the app. This tendency is also seen in food consumption tracking and recommendation apps, where the visual outlook, and a distinct and organised layout determine the marketability of the apps. In this regard, we have considered some aspects for evaluating the aesthetics of an app including -- layout consistency and readability, content resolution, visual appeal, and group targeting according to app content.

\subsubsection{General features}\label{sec:general}
General features (such as data sharing options, sign-up, etc.) are crucial for enhancing the user experience of food consumption tracking and recommendation apps. The log in or sign-up feature is important to preserve the users' food consumption history data in case of users changing devices. Features such as data export and share options are crucial for the users to use the data for other purposes (e.g., share with a nutritionist/dietitian). Sending regular notifications is also considered an important feature because this reminds users when to consume food. Furthermore, tutorial or on-boarding facilities are also considered a desirable feature since they demonstrate the  operations the app. The relevance of content customisation and the amount of visual information provided in apps have also been noted in recent years for boosting an app's user value. Therefore, they were also included in the general features. Moreover, a subscription package is also regarded as a useful factor because it can help support the development of a better user experience.

\subsubsection{Performance and efficiency}
One of the key features contributing to an app's acceptance to users is its efficiency and performance. Efficiency relates to, how fast an app functions and gives results on a device. The performance of an app includes battery life, device heating, and so on. However, the  performance metrics may differ depending on the mobile device hardware configuration. These measures play a crucial role in an app’s characterisation and have been included as a sub-scale for rating food consumption tracking and recommendation apps.

\subsubsection{Usability}

 The usability of mobile apps has become a significant issue because many software products currently running on smartphones previously ran on desktops and laptops~\citep{hussain2017user}. Users do not like apps that have a poor standard of usability and lack appropriate user-centered design. It is vital to test the usability of food consumption tracking and recommendation apps to identify whether they have sufficient characteristics to capture the interest of their target user groups. Nowadays, the attention of app users can be divided into two ways -- one is through interaction with the app, and the other is  with the environment~\citep{kallio2005usability}. Navigation and ease- of- use are also crucial indicators of app usefulness. The screen sequences in an app guide users through the various views, allowing them to receive the needed information from the app~\citep{georgieva2011evaluation}. Because of the differences in user behavior and user experience, the usability of an app in real life differs from the usability of an app in laboratory settings~\citep{kallio2005usability}. To asses the usability of the apps, we have focused on the following criteria: (i) the app can be used quickly and efficiently, (ii) the app's navigation activity is not disturbed, (iii) the gestural design and screen links (e.g., navigation panels buttons, arrows, etc.) are needed to be reconcilable throughout all the app pages, (iv) apps should provide an engaging experience by encouraging user input and providing feedback as appropriate.

\subsubsection{Functionality}\label{Specific functionalities}


In a food consumption tracking and recommendation app, the functionalities provided by the apps posses significant importance. The consideration of potential utility between two apps is determined by the integrated app functionalities. Several core functionalities are directly or indirectly involved with food consumption tracking and recommendations. These are food recognition, volume estimation, nutrition estimation, visualization of food consumption history, ability to add new food, and a food recommendation system. Table \ref{overall} summarizes the definition of the rating scores used to measure functionalities.

\begin{table}[!htbp]
    \centering
    \caption{App functionality measurement criteria and their ratings} 
    \label{overall}
    \resizebox{1\textwidth}{!}{
    \begin{tabular}{ lccccc}
\hline
Measurement criteria & Rating 5 & Rating 4 & Rating 3 & Rating 2 & Rating 1\\
\hline\hline
Food recognition & \makecell[l]{Food image and manually} & Food image & Barcode &  Manually & Doesn't support \\
Volume  estimation & \makecell[l]{Food image and manually} & Food image & Barcode &  Manually & Doesn't support  \\
Nutrition estimation & \makecell[l]{Food image and manually} & Food image & Barcode &  Manually & Doesn't support  \\
History visualization & \makecell[l]{More than 3 ways}& 3 ways & 2 ways & 1 way &  None \\
Food  recommendation & \makecell[l]{Based on nutrition consumption} & Calorie consumption &  Individual's preferences  & General healthy food & Doesn't provide \\
Allow to add new food item &  \makecell[l]{Automatically from users community} &  -- &  Manually &  --  &  Doesn't allow\\

\hline
 
 \hline
 
\end{tabular}
}
\end{table}


The key functionality we looked for is whether an app could recognize a food item. Novel food recognition systems for dietary evaluation have been enabled by the increasing use of smartphones and the advancement of artificial intelligence and computer vision technologies.
Various studies have already been conducted in this domain~\citep{zheng2017food, he2015dietcam,ravi2015real,NAYAK2020100297,chopra2021recent}. The first thing users need to do in food consumption tracking and recommendation apps is  track their food consumption. Thus, it is essential for an app to allow its users to input food details. Otherwise, the app would not be able to track users' food consumption history and provide recommendation accordingly. The functionality of adding food details(also called ``food recognition") can vary from automatic recognition from a food photo (taken by the user), scanning the barcode on a food packet, or manually entering information into the app or selecting from a database. Recognizing food from images is the most advanced technology in the field of food recognition. There are some recent studies on food recognition systems based on image recognition~\citep{ming2018food, aguilar2019regularized,KNEZ2020460,LIU2021193,mezgec2017nutrinet,10.1145/3391624}. 

Another functionality is volume estimation of users' consumed food. It is crucial to calculate food portion size or food volume to determine users' nutrition intake~\citep{min2019survey}. There are numerous studies that have been conducted to compute food volume size automatically~\citep{healthcare9121676,dehais2016two, fang2018single,yang2019image, tay2020current}. Furthermore, volume estimation is essential to compute food nutritional value, which is an importnat predictor of immunological responses~\citep{chandra1997nutrition}. Some studies have focused on estimating nutritional value from food images using artificial intelligence~\citep{KIRK2021104365,zhang2015snap, meyers2015im2calories,pouladzadeh2014measuring,BOLAND2019634,MICHEL2019194}. 

Studies~\citep{mezgec2017nutrinet, meyers2015im2calories,BOLAND2019634,LIU2021193} show that the most advanced technique to recognize food items, estimate their volume and  nutritional value is automatic detection from an image. Automatic detection is considered good for these three aspects. So, in our modified rating scale, a rating of 5 is given to those apps that can recognize food, estimate its volume and nutritional value automatically from image and also provide the flexibility to  users to make the apps recognize food items manually. In this rating scale, manual food recognition means users must manually enter the food item's name and search for it from the respective app databases. A rating 4 is given to those apps that can provide these functionalities through automatic image recognition. Furthermore, barcodes are already widely employed in industries and commercial sectors such as transportation, technology, food production and so on~\citep{sriram1996applications}. Hence, this feature was given a rating of 3 for food recognition where users' need to scan the barcode of a food packet to recognize it. But barcode scanning is not suitable for all types of food, as many food items do not come with a barcode (e.g., unpacked food). Besides, in the case of portion and nutrition estimation, the barcode scanning feature is also given a rating of 3. Moreover, barcode scanning has some drawbacks- foods are not consumed the way they are packed in a package. After being cooked, the nutritional value will change. It is impossible to accurately estimate food volume as user may not consume the exact amount of food present in the package. On the other hand, the barcode feature relieves users from manually inputting food data. Hence, a rating of 3 does justice to this feature. In the manual system, where the only way apps can support these functionalities is through user input, app were given a rating of 2. If an app is unable to support these functionalities, it is rated as 1.

Increasing a person's knowledge and understanding of their eating patterns and motives for eating may aid in beneficial dietary changes. Besides, data visualization may be useful in determining correlations between dietary and behavioral aspects~\citep{hingle2013collection}. That is why, visualization of consumption history is another significant criterion of measuring functionalities. It helps users to visualize their food or calorie consumption history in charts or graphs. This functionality is good for users to see what they are consuming and improve their food habits. If any user needs to show their food consumption habit to a dietitian, this functionality will be the greatest advantage. We rated an app 5 if it provides the user flexibility to see their food consumption history more than three ways, like yearly, monthly, weekly, daily and so on. A rating of 4 is given to those that have this feature in three ways (daily, monthly, weekly). Finally, the apps with a one-way visualization feature are rated 2, and those have with two way visualization are rated 3.

Food recommendation is an excellent way to suggest users eat healthy foods~\citep{10.1145/3418211}. A recommendation system uses information from a user's profile and compares it to come up with a list of relevant suggestions \citep{vivek2018machine}. Food recommendation is also mandatory when users want to know what they should consume as per their current health status. If an app can suggest appropriate food names, it helps immensely. For this reason, we considered the food recommendation option as another specific functionality of food consumption tracking and recommendation apps. 
For a user, it is very beneficial to know which type of food they need to consume to fulfill their bodies' nutritional needs. Hence, any app that can automatically suggest foods based on the nutritional is rated as 5, as nutritional components are the most important factors for a balanced diet and good health~\citep{elsweiler2015bringing}. In another work, the recommendation is made based on calorie count~\citep{ge2015health}. Calories are energy, and the number of calories tells us very little about a dish's nutritional content, both in terms of macro-nutrients (protein, carbohydrates, and fat) and micro-nutrients (vitamins, minerals, and phytonutrients like antioxidants). That is why,a rating of 4 is given to those apps that can recommend food to users based on calorie estimation. A rating of 3 was given to the apps that suggest food based on the users’ preferences. If a user prefers a meat item or fruit items, the app suggests food from those preferred domains and in this case, health issues are not considered. Finally, the apps that can suggest food items that are generally good for the human body and are not considering any specific factors, such as  recommending fruits, vegetables, milk, eggs, and so on, are rated a 2 for the food recommendation function. 

There is a wide variety of food worldwide. It is extremely challenging to create a database with all the food names with their nutritional values. Thus, there should be an option  for users to add new food items to the apps' food database. We determined that the  functionality of adding new food items to the database should be present in a food consumption tracking and recommendation app, as this option can  benefit future users of the apps. Apps that allow users to automatically add new food items  from their users' community discussion are rated as 5. The databases of these apps can become enriched with numerous entries of food items due to this feature. Apps that allow users to add food items manually, i.e., manually entering a food item's details in the device database, were rated a 3. We found many apps that do not allow the user to add new food items to their database, and we rated these apps 1 in this functionality category.



\subsubsection{Transparency}
Most mobile apps depend on social and personal information to work properly. Various businesses that profit from customized services commonly target this information~\citep{brug1995psychosocial}. Often, the app developers or publishers sell private information to third parties without the permission of the users, which violates users' privacy. The apps must follow strict and precise data protection and regulation laws, such as asking users if they consent to their private data being accessed. The apps should clearly state how and why users' data is being collected, even if the users are unaware of the direct effects of such acts. In the case of food consumption tracking and recommendation apps, constraints such as ``do not share private data", ``considering user consent in case of sharing", and ``verification of the developer" should be observed, which will assist users in determining whether the source of the app can be trusted or not. Furthermore, it is a matter of investigation to see if the software can meet the goals indicated in the store description.
 
 \subsubsection{Subjective quality}
 App subjective quality refers to the users' perspectives of the app~\citep{ashad2020mobile}. We used several metrics to assess the subjective quality of individual applications, including assessing personal app scores, preferring to pay for an app depending on its functionalities, preferring to recommend an app, and reviewing positive and negative feedback about the app. An overview of the app's offerings can be made by looking at the reactions of users who downloaded and used the application. However, this is a subjective viewpoint, and this method of assessing an app's performance prior to download is ineffective for apps with few or no user comments or ratings on the app stores. Nowadays, users tend to comment on the app store with more details and critical points,making it easier for new users to find useful apps easily. Therefore, the subjective quality of apps is an optional but valid criterion to find effective and preferable apps.
 
 \subsubsection{Perceived impact of app on users}
 The impact on a user's perception after using an app can be used to assess the app's potential. There are certain features like awareness, attitude and behavior changes, help-seeking attitude, and so on, to check this potentiality. It is essential to identify whether an app can spread awareness among the users or not. Our desired apps should be able to alert people about health issues or the impact of poor food habits. Another aspect is knowledge enhancing behavior. Apps should increase a users knowledge about food items. For example, the nutritional value of foods,positive impacts of foods on health and the body. Also, users may learn more about what food they need to avoid or the harmful effects of any food items by gathering knowledge from the app. The main impact of a food consumption tracking app is whether it can change users' attitudes toward improving their diet. The app can play a vital role in encouraging users to consume healthy food, and maintain good food routines. Furthermore, users' approaches to seeking health and food-related help can be perceived as another impact of app on users. Our study also assessed the impact of the app on users to understand the perceived impact of the applications on users, and whether these features were present in the applications. 
 
 
 \section{Results}\label{result}

\subsection{Inter-rater and intra-rater reliability}
Inter-rater reliability is a way of quantifying the level of agreement between two or more raters who rate an item (in the case, an app) independently based on a set of criteria~\citep{lange2011inter}. We used the intra-class correlation (ICC) method to assess inter-rater reliability. ICC is one of the most widely used statistics for evaluating inter-rater reliability if a study includes two or more raters~\citep{sawa2007interrater}. In our study, all apps were rated by the same three raters. Thus, we have used the ICC two-way mixed model as it is recommended when the raters are fixed and each of the apps is rated by all raters~\citep{koo2016guideline}. Depending on the 95\% confidence interim of the ICC estimation, values smaller than 0.5, within 0.5 and 0.75, within 0.75 and 0.9, and higher than 0.90 suggest poor, moderate, good, and excellent reliability, respectively~\citep{koo2016guideline}. 
The ICC score of our 80 apps was calculated as 0.90 (95\% CI ranging from 0.89 to 0.91), showing a good level of inter-rater reliability.

Intra-rater reliability is estimated to measure how consistent an individual is at measuring a set of criteria. This is a reliability estimation in which the same evaluation is performed by the same rater on more than one occasion.
To measure the intra-rater reliability of the three raters, we randomly selected three apps from our included list of 80 apps. The selected three apps were in three levels of quality (as per their overall rating score): low, average, and high. Those three apps were: ``Calorie Counter - MyNetDiary, Food Diary Tracker", ``Stupid Simple Marcos IIFYM Tracker", and ``Nutrition Tracker". The three raters reviewed these three apps twice in two months. All three raters showed a significant good level of intra-rater reliability between their two ratings; their two-way mixed ICC values were 0.89 (95\% CI 0.85--0.93), 0.8 (95\% CI 0.72--0.86), and 0.88 (95\% CI 0.83--0.92), respectively.

\subsection{Internal consistency of modified scale}
Internal consistency measures the degree of inter-relationships or homogeneity among the items on a test (in our case the questions/items used in a sub-scale/assessment criteria), such that the items are consistent with one another and measuring the same thing~\citep{christmann2006robust}. We have used Cronbach's alpha which is the most popular means of calculating internal consistency~\citep{cronbach1951coefficient}. Cronbach's alpha ($\alpha$) reliability coefficient indicates internal consistency that ranges between 0 and 1, with  $0.9\le\alpha$ as excellent, $0.8\le\alpha<0.9$ as good, $0.7\le\alpha<0.8$ as acceptable, $0.6\le\alpha<0.7$ as questionable, $0.5\le\alpha<0.6$ as poor, and $\alpha<0.5$ as unacceptable~\citep{gliem2003calculating}. The closer the value to 1 the higher the internal consistency.
We have randomly chosen three apps -- ``Weight Loss Coach \& Calorie Counter - Nutright", ``Foodzilla! Nutrition Assistant, Food Diary, Recipe" and ``Fitatu Calorie Counter - Free Weight Loss Tracker" to compute internal consistency. Table~\ref{tab:ic-result} reports the internal consistency of the sub-scales of our devised rating scale -- aesthetics, performance, usability, subjective quality, transparency and perceived impact. We excluded two sub-scales -- general and functionality, as their items are not meant to be collective measures of the construct.
The overall internal consistency of our modified scale was high at alpha 0.93, which is as regarded an excellent by prior studies~\citep{ursachi2015reliable}. 



\begin{table}[htb]
    \centering
    \caption{Internal consistency of the rating scale}
    \label{tab:ic-result}
    \begin{tabular}{lcc}
        \hline
         Sub-scale & Cronbach's alpha & Internal consistency \\
         \hline\hline
         Aesthetics & 0.94 & Excellent\\
         Performance & 0.78 & Acceptable\\
         Usability & 0.71 & Acceptable\\
         Subjective & 0.92 & Excellent\\
         Transparency & 0.76 & Acceptable\\
         Impact & 0.95 & Excellent\\
         \hline
         Overall & 0.93 & Excellent\\
         \hline
         
         \hline
    \end{tabular}
\end{table}

\subsection{Overall assessment of the apps}

The sub-scale ratings of all 80 apps with their mean and standard deviation are reported in Table~\ref{section}. The rating of each sub-scale is computed by taking the mean of the scores of all items in that sub-scale. The scores an app received in different sub-scales were used to calculate its overall mean (and standard deviation), indicating overall quality. 

As discussed in Section~\ref{sec:aesthetics}, we analyze various items such as layout, graphics, visual appeal, and appropriateness for the targeted audiences to measure the aesthetics of an app. In terms of layout, graphics, and appropriateness, more than 90\% of the apps are rated above 4 out of 5, and 81.25\% (65/80) of the apps are rated above 4 in visual appeal. In aesthetics, 18 apps (22.5\%) received the highest score, i.e., 5 out of 5 and the lowest score was 1.5 (``WAIE What Am I Eating - v2").

The general features sub-scale is measured using items such as social sharing, login/sign-up, data export, notifications, subscription, tutorial, and customization, as discussed in Section~\ref{sec:general}. In our reviewed apps, 70\% (56/80) do not have any social sharing features. Between 30\% to 40\% of the apps do not have a login or sign-up option, regular notifications, and any premium subscription. For tutorials or onboarding,  58.79\% (47/80) of the apps do not have these facilities to help users to operate the apps. 67.50\% (54/80) of the apps do not allow the user to export their data. One-fifth of the apps (16/80) do not have the customization feature. In this sub-scale, the app ``Health Mate" received the highest score (4.86) and 5 apps (6.25\%) rated the lowest score of 1.

While rating the apps, we found that most of the apps scored 4 to 5 the in performance sub-scale, which means they are responsive, components are working well, the apps do not crash, and battery power and memory consumption are  reasonable.
The app ``Nutrition Tracker" is the only app to score 3 in this sub-scale as it had component and feature issues.

In the usability sub-scale, 80\% of the apps(64/80) were very easy to use, 77.5\% (62/80) had high navigational accuracy, 72.5\% (58/80) featured a very good quality of gestural design, and 63.75\% (51/80) were rated high in interactivity and user feedback, i.e., scored 5 out of 5 in all these areas. Overall, most of the apps scored high in this sub-scale -- 83.75\% (67/80) of the apps scored between 4 and 5, and the rest, i.e., 16.25\% (13/80) scored between 2.5 and 3.75. The lowest rated apps in this sub-scale were ``FoodImage" and ``WAIE What AM I Eating - v2", which both scored 2.5.

Functionality is an essential sub-scale in our rating scale as it measures the extent (not at all (1), manually (2) to fully automatically (5)) to which an app supports food recognition, food volume estimation, nutritional value estimation, food consumption history/pattern visualization, food recommendations, and the ability to add new food information in the app. In our reviewed apps, 78 out of 80 apps (97.5\%) scored below 3 (i.e., average). The app ``Foodvisor: Calorie Counter, Food Diary \& Diet Plan" (3.67) is one of the two apps that scored three and above. This result indicates that although there are many apps in the market, and significant advances in artificial intelligence and image processing technologies, there is still a lack of smart food computing apps with automatic desired features. 

The transparency sub-scale is measured based on an: app's description in the app store, it is credibility (from a legitimate source), evidence, goals, and policy in accessing and sharing user data. Eleven out of 80 apps (13.75\%) scored between 4 and 5. Among them four apps ``Calorie Counter - MyNetDiary, Food Diary Tracker", ``Calorie Counter - MyFitnessPal", ``Calorie Counter by Lose It! for Diet \& Weight Loss", and ``Lifesum - Diet Plan, Macro Calculator \& Food Diary" scored the same highest score of 4.40. The two apps that received the lowest score (1.60) were ``DietYuk" and ``WAIE What Am I Eating - v2".

The subjective quality sub-scale is measured based on an individual's willingness to use, recommend, and pay for the apps, and the overall star rating given by the individual (in this case, the rater). In this sub-scale, only 8 out of 80 apps (10\%) scored 4 and higher. The app ``WW Weight Watchers Reimagined" scored the highest score of 4.5. The two apps that scored 1 (the lowest) were ``DietYuk" and ``WAIE What Am I Eating - v2". 

The perceived impact on user sub-scale measures the effectiveness of an app in changing users' attitudes toward a balanced diet and healthy life. This has been evaluated based on whether the app provides a diet plan considering an individual's eating habits, community or forum to share information, seek help, etc. Only 7 out of 80 apps (8.75\%) scored between 4 and 5; the only app that scored 5 is ``WW Weight Watchers Reimagined". On the other hand, five apps (6.25\%) scored 1 in this sub-scale. 

In addition, we found some other useful features while reviewing those apps. Tracking weight, tracking exercise and steps, and tracking water consumption are the most common features. The apps ``Carb Manager: Keto Diet Tracker \& Macros Counter", ``My Healthy Plate", ``MyDietDaily - Diet Watchers, Smart Weight Loss", and ``Keto Manager: Keto Diet Tracker \& Carb Counter App", could recognize food items from voice commands  as they had a voice recognition feature. Some apps, for example, ``Calorie Counter - MyNetDiary, Food Diary Tracker", ``Carb Manager: Keto Diet Tracker \& Macros Counter", ``Health Mate - Calorie Counter \& Weight Loss App", and ``Health \& Fitness Tracker with Calorie Counter" had some unique features like tracking sleep schedule, blood pressure, and heart rate. Some apps had a large database of recipes that the users liked. Many apps could sync with fitness apps like Fitbit, Nokia Health, Samsung Health, Misfit trackers, Garmin trackers, Withings scales, Google Fit, and Healthkit. A few apps offered menus from various restaurants'. 


{\small\tabcolsep=3pt  
\renewcommand{\arraystretch}{.83}
\begin{longtable}
{p{6cm}p{.8cm}p{.8cm}p{.8cm}p{.8cm}p{1cm}p{1cm}p{.8cm}p{.8cm}p{1.4cm}}
\caption{Assessment scores for food consumption tracking and recommendation apps}
\label{section}\\
\hline 
App name & \makecell[l]{Aesth\\etics} & \makecell[l]{Gene\\ral} & \makecell[l]{Perfor\\mance} & \makecell[l]{Usabi\\lity} & \makecell[l]{Function\\ality} & \makecell[l]{Transpa\\rency} & \makecell[l]{Imp\\act} & \makecell[l]{Subje\\ctive}  & \makecell[c]{Mean\\(Std Dev)}\\ \hline\hline
\endfirsthead
 \caption{Assessment scores for food consumption tracking and recommendation apps (continued)} \\
\hline 
App name & \makecell[l]{Aesth\\etics} & \makecell[l]{Gene\\ral} & \makecell[l]{Perfor\\mance} & \makecell[l]{Usabi\\lity} & \makecell[l]{Function\\ality} & \makecell[l]{Transpa\\rency} & \makecell[l]{Imp\\act} & \makecell[l]{Subje\\ctive}  & \makecell[c]{Mean\\(Std Dev)}\\ 
\hline
\endhead
\hline 
\endfoot
\hline 
\endlastfoot

 \hline
 
 \footnotesize WW Weight Watchers Reimagined & 4.75 & 4.28 & 5.00	& 4.50& 2.83 & 4.20& 5.00 & 4.50 &4.37 (0.75)\\

\footnotesize Calorie Counter - MyFitnessPal &4.75&	4.43& 5.00	&4.75&	2.50& 4.40&	4.17&4.25&4.29 (0.83)\\

\footnotesize Calorie Counter by Lose It! for Diet \& Weight Loss&4.50 & 4.14 & 5.00	&4.75	&2.67	&4.40 & 4.00 &3.75&4.21 (0.76)\\

\footnotesize Calorie Counter by FatSecret	&4.75	&4.67	&4.83	&5.00	&2.50	&4.20	&3.33	&3.00 &4.18 (0.93)\\

\footnotesize Calorie Counter - MyNetDiary, Food Diary Tracker&	5.00	&3.57	&5.00	&4.75	&2.33	&4.40	&3.83	&3.75&4.13 (0.97)\\

\footnotesize Calorie, Carb \& Fat Counter &4.75	&3.14	&5.00	&5.00	&2.50	&4.40	&3.83	&3.75 &4.09 (0.97)\\

\footnotesize Calorie Counter - EasyFit free	&5.00	&4.43	&5.00	&5.00	&2.33	&3.60	&3.00	&4.00 &4.05 (1.09)\\

\footnotesize Calorie Counter +	&4.75	&4.57	&4.67	&4.75	&2.50	&3.60	&3.33	& 3.00 & 4.02 (0.89)\\

\footnotesize MyDietDaily - Diet Watchers, Smart Weight Loss	&4.75	&3.29	&5.00	&5.00	&2.50	&3.80	&3.83	&3.00 & 4.02 (0.95)\\

\footnotesize Lifesum - Diet Plan, Macro Calculator \& Food Diary	&5.00	&3.14	&5.00	&4.00	&2.50	&4.40	&4.00	&3.75 &4.01 (0.93)\\

\footnotesize YAZIO Calorie Counter \& Intermittent Fasting App	&4.75	&3.71	&5.00	&5.00	&2.33	&3.60	&3.67	&3.50 &4.01 (0.98)\\

\footnotesize MyPlate Calorie Tracker	&4.25	&3.29	&5.00	&5.00	&2.83	&4.20	&3.33	&3.50 &3.99 (0.46)\\

\footnotesize Fitstyle - Diets \& Workouts	&5.00	&3.29	&5.00	&5.00	&2.50	&3.60	&3.33	&4.25 &3.96 (1.02)\\

\footnotesize TrackEats	&5.00	&4.29	&5.00	&5.00	&2.67	&2.40	&3.33	&3.75 &3.96 (1.14)\\

\footnotesize Fitatu Calorie Counter - Free Weight Loss Tracker	&4.75	&4.14	&4.83	&4.75	&2.67	&2.60	&3.67	&3.75 &3.92 (0.97)\\

\footnotesize Carb Manager: Keto Diet Tracker \& Macros Counter 	&4.75	&4.00	&5.00	&4.00	&2.67	&3.60	&3.33	&3.50 &
3.91 (0.80)\\

\footnotesize Calorie Calculator - EatRytte	&4.75	&3.43	&5.00	&5.00	&2.33	&3.60	&3.00	&3.50 &3.87 (1.05)\\

\footnotesize Total Keto Diet: Low Carb Recipes \& Keto Meal Plan	&5.00	&3.00	&5.00	&5.00	&2.33	&3.60	&3.17	&3.75 &3.87 (1.11)\\

\footnotesize Calorie counter	&5.00	&4.29	&5.00	&5.00	&2.50	&2.40	&2.50	&3.75 &3.81 (1.28)\\

\footnotesize Foodzilla! Nutrition Assistant, Food Diary, Recipe	&5.00	&1.86	&4.83	&5.00	&3.17	&3.40	&3.33	&3.75 &3.80 (1.18)\\

\footnotesize MunchLog Calorie Counter \& Meal Planner (BETA)	&5.00	&3.00	&5.00	&5.00	&2.50	&2.40	&3.67	&3.50 &3.80 (1.20)\\

\footnotesize Health \& Fitness Tracker with Calorie Counter	&4.75	&4.14	&4.83	&4.50	&2.33	&2.80	&3.17	&3.00 &3.79 (1.01)\\

\footnotesize Yamfit - calorie counter, diet and meal planner	&4.75	&3.57	&5.00	&5.00	&2.00	&3.40	&2.83	&3.50 &3.79 (1.17)\\

\footnotesize Cronometer – Nutrition Tracker	&4.25	&3.57	&4.83	&4.25	&2.83	&4.20	&2.50	&3.00 &3.78 (0.85)\\

\footnotesize Health Mate - Calorie Counter \& Weight Loss App	&4.50	&4.86	&5.00	&4.50	&2.00	&2.40	&3.17	&3.50 &3.78 (1.23)\\

\footnotesize Bitesnap: Photo Food Tracker and Calorie Counter	&4.25	&3.43	&5.00	&5.00	&3.00	&3.40	&2.33	&3.50&3.77 (1.01)\\

\footnotesize Foodvisor: Calorie Counter, Food Diary \& Diet Plan	&4.00	&2.43	&4.83	&5.00	&3.67	&3.20	&3.17	&4.00 &3.76 (0.93)\\

\footnotesize Healthy Diet - Best Diet Plan, Calorie Counter	&4.75	&2.57	&5.00	&5.00	&2.33	&3.40	&3.00	&3.50 &3.72 (1.17)\\

\footnotesize Track - Calorie Counter	&4.25	&2.71	&5.00	&5.00	&2.33	&3.60	&3.17	&3.25&3.72 (1.07)\\

\footnotesize Private Calorie Counter - OmNom Notes	&4.75 &3.00	&5.00	&5.00	&2.17	&3.60	&2.50	&3.00 &3.72 (1.20)\\

\footnotesize Keto.app - Keto diet tracker &5.00	&2.14	&5.00	&5.00	&2.20	&3.60	&3.00	&4.00 &3.71 (1.30)\\

\footnotesize AI Nutrition Tracker: Macro Diet \& Calorie Counter	&4.50	&3.29	&4.67	&4.75	&2.17	&3.40	&3.17	&3.00 &3.71 (0.96)\\

\footnotesize Keto Manager: Keto Diet Tracker \& Carb Counter	&4.25	&3.71	&5.00	&5.00	&2.17	&2.60	&3.17	&3.00 &3.70 (1.11)\\

\footnotesize Calories tracker, diet diary \& lose weight	&5.00	&2.43	&5.00	&5.00	&2.40	&2.80	&3.17	&3.50 &3.69 (1.25)\\

\footnotesize Calorie Counter \& Diet Tracking	&5.00	&3.29	&5.00	&4.00	&2.33	&2.20	&4.00	&3.75 &3.69 (1.14)\\

\footnotesize iEatBetter: Food Diary	&4.00	&3.14	&4.83	&4.25	&2.17	&4.20	&3.17	&3.00 & 3.68 (0.90)\\

\footnotesize FitGenie: Macro \& Food Tracking	&4.75	&2.43	&5.00	&5.00	&2.33	&3.40	&2.83	&3.25 &3.68 (1.12)\\

\footnotesize FoodTracker: Calorie Counter	&5.00	&2.29	&5.00	&3.75	&2.33	&3.40	&4.00	&4.00 &3.68 (1.11)\\

\footnotesize Dietary Calorie Counter	&4.75	&2.86	&5.00	&5.00	&2.33	&2.60	&3.17	&3.00 &3.67 (1.19)\\

\footnotesize Calorie Counter, Nutrition Diary \& Diet Plan	&4.50	&3.29	&5.00	&5.00	&2.33	&2.80	&2.67	&3.50 &3.66 (1.15)\\

\footnotesize Calorie Counter - Fddb Extender	&4.75	&2.57	&5.00	&5.00	&2.33	&3.40	&2.50	&3.00 &3.65 (1.23)\\

\footnotesize MacrosFirst - Macro tracking made easy &5.00	&3.29	&5.00	&5.00	&2.33	&2.40	&2.50	&3.50 &3.65 (1.30)\\

\footnotesize Calorie Counter, Carb Manager \& Keto by Freshbit	&4.50	&3.43	&4.67	&4.75	&2.17	&2.80 &3.17	&3.50 &3.64 (1.01)\\

\footnotesize FoodPrint™ - Nutrition Tracker &5.00	&2.43	&5.00	&5.00	&2.00	&3.40	&2.50	&2.50 &3.62 (1.36)\\

\footnotesize Diety - Diet Plan, Calorie Counter, Weight Loss	&4.00	&2.57	&5.00	&5.00	&2.33	&3.20	&3.17	&2.75 &3.61 (1.09)\\

\footnotesize Stupid Simple Macros IIFYM Tracker	&4.00	&3.43	&4.83	&5.00	&1.67	&3.40	&2.67	&2.75 &3.57 (1.17)\\

\footnotesize MealLogger-Photo Food Journal	&4.00	&3.29	&5.00	&4.50	&2.00	&4.20	&2.00	&3.00 &3.57 (1.19)\\

\footnotesize Weight Loss Coach \& Calorie Counter - Nutright	&4.50	&3.00	&5.00	&5.00	&1.67	&2.60	&3.17	&4.00 &3.56 (1.29)\\

\footnotesize Calorie Counter - Nutrition \& Healthy Diet plan	&4.00	&4.29	&5.00	&4.25	&2.33	&2.20	&2.67	&2.75 &3.53 (1.11)\\

\footnotesize Hol: Weight Loss Calorie Counter Nutrition Tracker	&4.50	&3.00	&5.00	&4.75	&2.00	&2.40	&3.00	&2.75 &3.52 (1.21)\\

\footnotesize Macros - Calorie Counter \& Meal Planner	&4.00	&3.29	&5.00	&4.75	&2.33	&2.20	&3.00	&3.00 &3.51 (1.11)\\

\footnotesize See How You Eat Food Diary App	&4.25	&4.43	&4.83	&4.75	&1.17	&3.40	&1.33	&1.75 &3.45 (1.57)\\

\footnotesize Cornflakes - Calorie Counter - Diet and Fitness	&4.50	&3.57	&4.83	&3.75	&2.00	&3.00	&2.50 	&3.50 &3.45 (1.02)\\

\footnotesize Healthy Habeats 2.0	&5.00	&1.57	&5.00	&4.50	&1.67	&3.60	&2.50	&3.50 &3.41 (1.50)\\

\footnotesize Nutritionist+	&4.25	&3.00	&5.00	&4.75	&2.33	&2.20	&2.17	&2.75 &3.39 (1.25)\\

\footnotesize Lilbite Food Tracker – Calorie Counter \& IIFYM	&5.00	&1.57	&5.00	&5.00	&2.00	&2.20	&2.83	&3.00 &3.37 (1.57)\\

\footnotesize KetoDiet: Keto Diet App Tracker, Planner\& Recipes	&4.50	&2.57	&5.00	&3.75	&2.00	&2.20	&3.00	&2.00 &3.29 (1.16)\\

\footnotesize Meal Tracker	&4.25	&2.71	&5.00	&4.00	&2.00	&2.60	&2.17	&2.75 &3.25 (1.16)\\

\footnotesize Nutrition Tracker	&3.75	&1.86	&4.50	&4.75	&2.17	&3.20	&2.17	&2.25 &3.20 (1.17)\\

\footnotesize Keto diet tracker and macros calculator	&4.25	&1.00	&4.83	&4.00	&2.33	&3.80	&2.17	&2.25 &3.20 (1.38)\\

\footnotesize Food and Weight Tracker Lite - Calorie Counter	&3.50	&2.29	&4.83	&4.00	&2.00	&3.20	&2.17	&2.50 &3.14 (1.05)\\

\footnotesize Calorie Calculator	&4.00	&3.14	&5.00	&3.75	&1.67	&1.8	&2.17	&2.50 &3.08 (1.25)\\

\footnotesize Meal Tracker(Calorie Tracker, Weight Loss)	&4.25 &1.86	&4.67	&4.00	&2.17	&2.40	&2.17	&2.75 &3.07 (1.18)\\

\footnotesize My Healthy Plate	&4.00	&1.71  &5.00	&4.00	&1.67	&2.60	&2.33	&2.75 &3.04 (1.29)\\

\footnotesize Feather Weight	&5.00	&1.57	&4.83	&4.00	&1.50	&2.40	&2.00	&2.25 &3.04 (1.53)\\

\footnotesize Diet Clock	&4.00	&1.57	&5.00	&4.00	&2.00	&2.40	&2.17	&2.75 &3.02 (1.29)\\

\footnotesize Food Calorie Calculator	&4.00	&1.57	&4.67	&3.50	&1.67	&3.60	&2.00	&2.00 &3.00 (1.24)\\

\footnotesize Calorie Tips	&4.25	&1.57	&5.00	&5.00	&1.15	&1.80	&2.00	&1.25 &2.97 (1.70)\\

\footnotesize Simple Diet Diary	&3.75	&2.86	&5.00	&3.25	&2.17	&2.00	&1.50	&1.75 &2.93 (1.19)\\

\footnotesize Dr. Greger's Daily Dozen	&3.50	&1.71	&4.83	&3.50	&1.50	&3.40	&2.00	&2.25 &2.92 (1.21)\\

\footnotesize Calorie Counter - Food \& Diet Tracker	&4.00	&1.00	&4.83	&4.00	&1.67	&2.20	&2.67	&2.25 &2.91 (1.40)\\

\footnotesize Daily Calories (offline)	&3.75	&1.00	&5.00	&3.75	&1.67	&3.20	&2.00	&2.75 &2.91 (1.40)\\

\footnotesize Food Diary	&2.75	&2.85	&4.83	&4.00	&1.17	&2.60 &1.00	&1.75 &2.74 (1.38)\\

\footnotesize NutritionCalculator	&3.50	&1.57	&5.00	&4.00	&1.67	&1.80	&1.67	&2.00 &2.74 (1.40)\\

\footnotesize Food-Tracker (Privacy Friendly)	&3.00	&1.43	&4.67	&3.25	&2.00	&2.20	&1.67	&2.00 &2.60 (1.13)\\

\footnotesize Diet Tracker Food Scanner	&3.25	&1.00	&5.00	&3.25	&1.83	&1.80	&2.00	&2.00 &2.59 (1.34)\\

\footnotesize FoodImage	&2.50	&2.43	&4.67	&2.50	&1.50	&2.60	&1.00	&1.75 &2.46 (1.15)\\

\footnotesize DietYuk	&3.25	&1.00	&5.00	&4.00	&1.33	&1.60	&1.00	&1.00 &2.45 (1.62)\\

\footnotesize Calorie counter	&3.25	&1.00	&5.00	&3.25	&1.00	&2.40	&1.00	&1.25 &2.41 (1.53)\\

\footnotesize WAIE What Am I Eating - v2	&1.50	&1.57	&4.50	&2.50	&1.13	&1.60 	&1.00	&1.00 &1.97 (1.21)\\

\end{longtable}
 }

Overall, seven out of 80 apps (8.75\%) scored mean values between 4 and 5. However, none of those seven apps scored highest in the functionality sub-scale. The app that scored highest in functionality (``Foodvisor: Calorie Counter, Food Diary \& Diet Plan") received an overall score of 3.79. Out of these seven top scoring apps, ``WW Weight Watchers Reimagined" received the highest score of 4.38. However, this app scored lowest in functionality among its sub-scale scores. Of all the apps, 78.75\% (63/80) scored above average (between 3 and 5) and only one (``WAIE What Am I Eating - v2") scored below 2.

\begin{figure}[!htb]
\centering
\resizebox{8cm}{!}{
\begin{tikzpicture}
    \begin{axis}[
        symbolic x coords={Aesthetics, General, Performance, Usability, Functionality, Transparency, Impact, Subjective, Total mean},
        x tick label style={font=\huge, rotate=45, anchor=east},
        ybar=-0.8cm,
        ymin=0,
        xtick distance=1,
        bar width=0.8cm,
        ytick={0,0.5,1,1.5,2,2.5,3,3.5,4,4.5,5},
        ylabel={Rating},  
        xlabel={Sub-scales},
        label style={font=\huge},
        y tick label style={font=\LARGE},
    ]
        \addplot+ [draw = blue,
       fill=blue!50,
            error bars/.cd,
                y dir=both,
                y explicit,
        ] coordinates {
            (Aesthetics,4.34) +-(0,0.6647)
            (General,2.8585) +- (0,1.0253)
            (Performance,4.9246) +- (0,0.1268)
            (Usability,4.4625) +- (0,0.6328)
            (Functionality,2.1547) +- (0,0.4777)
            (Subjective,2.99) +- (0,0.791)
            (Transparency,3.0475) +- (0,0.7698)
            (Impact,2.7443) +- (0,0.8206)
        };
        \addplot [draw =orange,
        fill = orange!50,
            error bars/.cd,
                y dir=both,
                y explicit,
        ] coordinates {
        (Total mean,3.4404) +- (0,0.5149)
        };
    \end{axis}
\end{tikzpicture}
}
\caption{Sub-scale specific ratings and overall rating} \label{fig:my_label}
\end{figure}
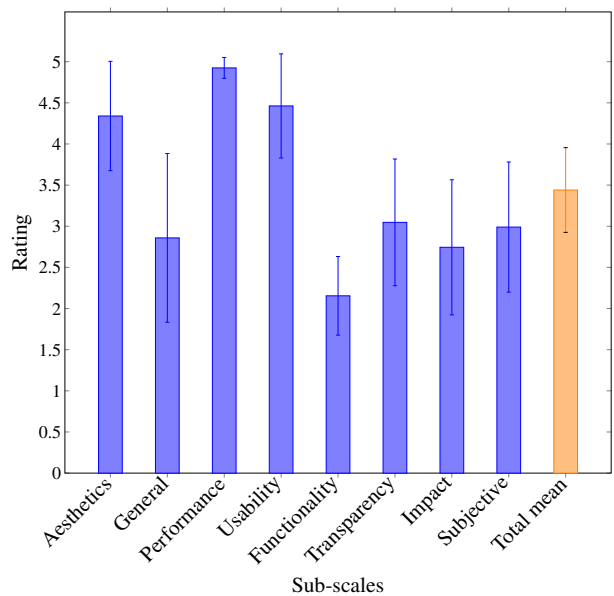

Figure~\ref{fig:my_label} shows the sub-scale specific scores and the total mean score of all the apps. The total mean score of all the apps was 3.44 out of 5 with a 95\% CI ranging from 3.33 to 3.55. Significant discrepancies were found within the sub-scales, most notably functionality, perceived impact, and general features, which received the lowest mean scores of 2.15, 2.74 and 2.86, respectively. Performance, usability, and aesthetics, on the other hand, received the highest mean scores of 4.92, 4.46 and 4.34, respectively. Transparency and subjective quality were two other sub-scales that received average mean scores of 3.05 and 2.99, respectively. In summary, the apps are highly lacking in functionality and very good in performance.

\subsection{Analysis of app store ratings and our measured ratings}
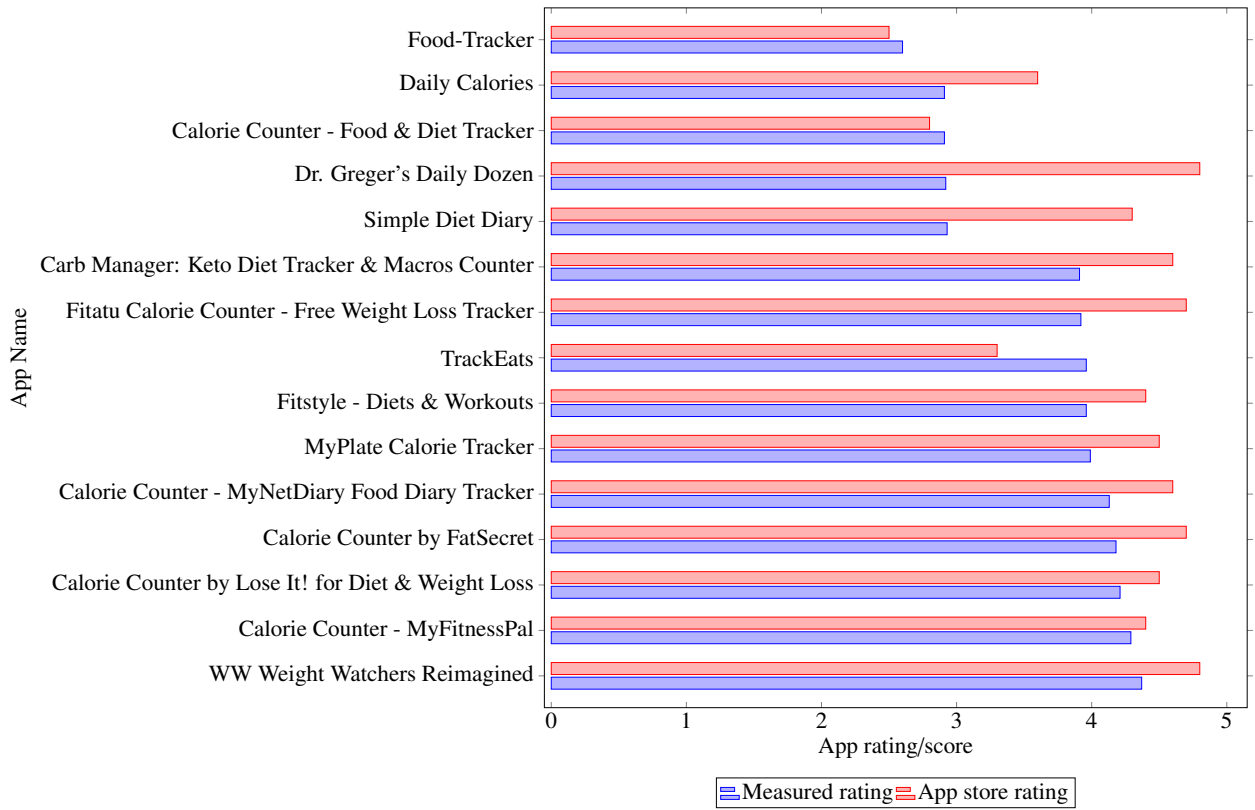
\begin{figure}[!htb]
\centering
\resizebox{1\textwidth}{!}{
\begin{tikzpicture}
\begin{axis}[
    xbar,
    bar width= 9pt,
    ylabel={App Name},  
    xlabel={App rating/score},
   label style={font=\LARGE},
    legend style={at={(0.5,-0.1),font=\LARGE},
      anchor=north,legend columns=-1},
    symbolic y coords={WW Weight Watchers Reimagined, Calorie Counter - MyFitnessPal, Calorie Counter by Lose It! for Diet \& Weight Loss, Calorie Counter by FatSecret, Calorie Counter - MyNetDiary Food Diary Tracker, MyPlate Calorie Tracker, Fitstyle - Diets \& Workouts, TrackEats, Fitatu Calorie Counter - Free Weight Loss Tracker, Carb Manager: Keto Diet Tracker \& Macros Counter, Simple Diet Diary, Dr. Greger's Daily Dozen, Calorie Counter - Food \& Diet Tracker, Daily Calories, Food-Tracker},
    ytick=data,
    xtick={0,1,2,3,4,5},
    y=1.2cm,
    xmin=0,
    xmax=5.1,
    enlarge y limits  = 0.05,
    enlarge x limits  = 0.01,
    scaled ticks=false,
    x tick label style={font=\LARGE},
    y tick label style={font=\LARGE},
    ytick distance=1,
    ]

\addplot coordinates {(4.37,WW Weight Watchers Reimagined) (4.29,Calorie Counter - MyFitnessPal) (4.21,Calorie Counter by Lose It! for Diet \& Weight Loss) (4.18,Calorie Counter by FatSecret) (4.13,Calorie Counter - MyNetDiary Food Diary Tracker) (3.99,MyPlate Calorie Tracker) (3.96,Fitstyle - Diets \& Workouts) (3.96,TrackEats) (3.92,Fitatu Calorie Counter - Free Weight Loss Tracker) (3.91,Carb Manager: Keto Diet Tracker \& Macros Counter) (2.93,Simple Diet Diary) (2.92,Dr. Greger's Daily Dozen) (2.91,Calorie Counter - Food \& Diet Tracker) (2.91,Daily Calories)(2.6,Food-Tracker)};

 \addplot coordinates {(4.8,WW Weight Watchers Reimagined) (4.4,Calorie Counter - MyFitnessPal) (4.5,Calorie Counter by Lose It! for Diet \& Weight Loss) (4.7,Calorie Counter by FatSecret) (4.6,Calorie Counter - MyNetDiary Food Diary Tracker) (4.5,MyPlate Calorie Tracker) (4.4,Fitstyle - Diets \& Workouts) (3.3,TrackEats) (4.7,Fitatu Calorie Counter - Free Weight Loss Tracker) (4.6,Carb Manager: Keto Diet Tracker \& Macros Counter) (4.3,Simple Diet Diary) (4.8,Dr. Greger's Daily Dozen) (2.8,Calorie Counter - Food \& Diet Tracker) (3.6,Daily Calories) (2.5,Food-Tracker)};

\legend{Measured rating, App store rating}
\end{axis}
\end{tikzpicture}}
\caption{Consistency between app store rating and measured rating}
\label{fig4comparison between app store and measured rating scale}
\end{figure}

The pearson correlation between app store ratings and our measured ratings is 0.46, which is considered a moderate strength \citep{liang2019effective} between these two values. Besides, our examined apps' standard deviation between the average app store ratings and our measured ratings was 0.49. Given that our rating scale's score is an aggregated mean of the different sub-scale ratings required to determine the consistency and parameters of food consumption tracking applications, this variance is not too low. 
Figure~\ref{fig4comparison between app store and measured rating scale} shows app store ratings and our measured ratings of the selected apps. In Figure~\ref{fig4comparison between app store and measured rating scale}, we have reported 15 apps out of 80. Here, we have considered the apps with 100+ user ratings in the app store, and randomly selected five apps from each measured rating range (4 to 5, 3 to below 4, 2 to below 3 and 1 to below 2). Figure~\ref{fig4comparison between app store and measured rating scale} shows a clear consistency between the app store ratings and our measured ratings.

\subsection{Assessment of functionality sub-scale}
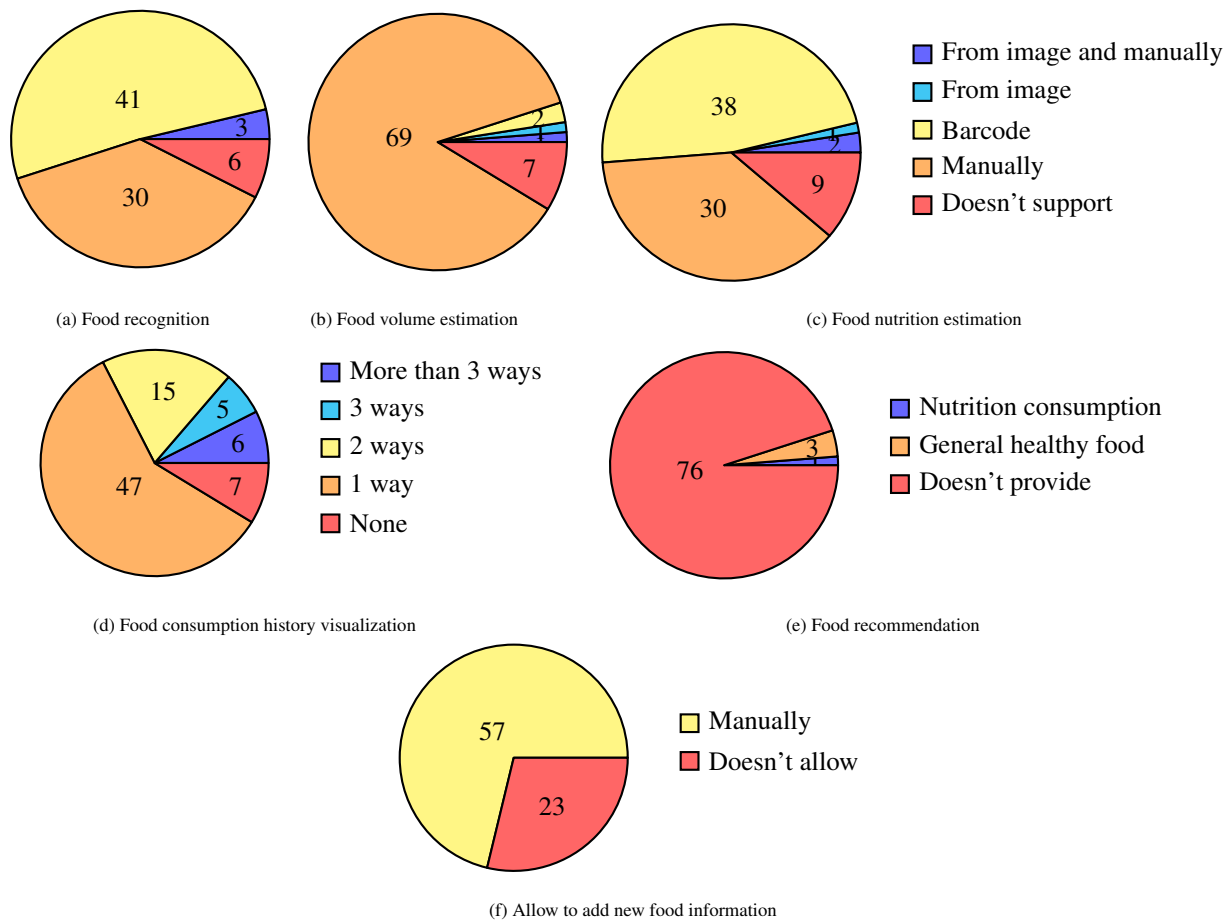
\begin{figure}[!htb]
\centering
\begin{subfigure}[b]{0.2\textwidth}
\centering
\begin{tikzpicture}
  \pie[sum=auto, radius=1.7,color={blue!60, yellow!60, orange!60, red!60}]{3/, 41/, 30/, 6/}
  \end{tikzpicture}
  \caption{Food recognition}
\end{subfigure}
\hfill
\begin{subfigure}[b]{0.2\textwidth}
  \centering
  \begin{tikzpicture}
  \pie[sum=auto,  radius=1.7,color={blue!60, cyan!60,  yellow!60, orange!60, red!60}]{1/,  1/, 2/, 69/, 7/}
  \end{tikzpicture}
  \caption{Food volume estimation}
\end{subfigure}
\hfill
\begin{subfigure}[b]{0.55\textwidth}
\centering
  \begin{tikzpicture}
  \tikzset{lines/.style={draw=white},}
  \pie[text=legend, sum=auto, radius=1.7,color={blue!60, cyan!60,  yellow!60, orange!60, red!60}]{2/From image and manually,  1/From image, 38/Barcode, 30/Manually, 9/Doesn't support}

 \end{tikzpicture}
\caption{Food nutrition estimation}
\end{subfigure}

\hfill
\begin{subfigure}[b]{.37\textwidth}
\centering
\begin{tikzpicture}
  \pie[text=legend, sum=auto, radius=1.5,color={blue!60, cyan!60,  yellow!60, orange!60, red!60}]{6/More than 3 ways,  5/3 ways, 15/2 ways, 47/1 way, 7/None}
  \end{tikzpicture}
  \caption{Food consumption history visualization}
\end{subfigure}
\hfill
\begin{subfigure}[b]{.6\textwidth}
\centering
\begin{tikzpicture}
  \pie[text=legend, sum=auto, radius=1.5,color={blue!60, orange!60, red!60}]{1/Nutrition consumption,  3/General healthy food, 76/Doesn't provide}
  \end{tikzpicture}
  \caption{Food recommendation}
\end{subfigure}
\hfill
\begin{subfigure}[b]{.4\textwidth}
\centering
\begin{tikzpicture}
  \pie[text=legend, sum=auto, radius=1.5,color={ yellow!60, red!60}]{57/Manually,  23/Doesn't allow}
  \end{tikzpicture}
  \caption{Allow to add new food information}
\end{subfigure}

\caption{Results of functionality assessment using six measurement criteria}
\label{Representation of specific functionality rating ratio}
\end{figure}

We have identified six criteria for measuring the functionality of the food consumption tracking and recommendation apps -- food recognition, food portion computing, nutritional value estimation, history visualization, food recommendation, and the ability to add food into their databases, as discussed in Section~\ref{Specific functionalities}. 
 Figure~\ref{Representation of specific functionality rating ratio} presents the results of our functionality assessment of 80 reviewed apps based on the six measurement criteria. 

Recognizing the food items from the recorded entries of the users is one of the key criteria for assessing the functionality of food tracking and recommendation apps. Among 80 reviewed apps, three could recognize food items both automatically and manually. These apps are -- ``Foodzilla! Nutrition Assistant, Food Diary, Recipe", ``Bitesnap: Photo Food Tracker \& Calorie Counter", and ``Foodvisor: Calorie Counter, Food Diary \& Diet Plan". Forty-one out of the 80 apps (51.25\%) have a barcode feature to recognize food items. Thirty out of the 80 apps (37.5\%) need manual input from users, as they do not have any automatic food recognition features. 

  
Another criterion that we have tested in the apps is whether they can compute food volume. Most of the apps cannot compute volume directly from an image, and thus, users have to enter the food volume manually. Among all the reviewed apps, 87.5\% (69/80) had the feature of computing food volume manually. We found only one app ``Foodzilla! Nutrition Assistant, Food Diary, Recipe" that provides both a manual and an automatic food volume computation feature. The app ``Foodvisor: Calorie Counter, Food Diary \& Diet Plan" can estimate food volume from images automatically. Moreover, two of the apps can compute volume using a barcode scanner -- ``Calories tracker, diet diary \& lose weight" and ``Calorie Counter - MyNetDiary, Food Diary Tracker".

We have also searched for the nutrition value estimation functionality in the reviewed apps. ``Foodzilla! Nutrition Assistant, Food Diary, Recipe", and ``Bitesnap: Photo Food Tracker and Calorie Counter" are the apps in which users can estimate nutrition value manually and automatically. Besides, we have found only one app, ``Foodvisor: Calorie Counter, Food Diary \& Diet Plan", that can compute food nutritional value through images. Thirty-eight apps can estimate the nutritional value of 100 gm of food from their database using barcode scanning. In 37.5\% of apps(30/80), users are required to input nutritional values manually.


Another major criterion that we have searched in the reviewed apps is food consumption history visualization. Most of the apps can show the records of user's daily consumption. Some apps can visualize weekly and monthly food consumption, and some can show yearly food consumption history. Only a few apps can visualize any records between a given time range defined by the users. More than half of the apps, 58.75\% (47/80), can show the history in only one of the ways mentioned above. Fifteen out of the 80 apps (18.75\%) can show the consumption history in two ways, 6.25\% (5/80) can portray the consumption record in three ways, and 7.5\% (6/80) can visualize the records in more than three ways (e.g, daily, weekly, monthly, yearly). 

Recommending foods according to the user’s requirements is one of the most important criteria of food consumption tracking and recommendation apps. However, in our study, we have found only one app that can recommend food based on the nutritional consumption of the user, named ``WW Weight Watchers Reimagined". Only 3.75\% of the apps (3/80) recommend foods that are good for health in general, but they do not consider any individual user requirements the consumption history of nutrition or calories. These three apps are-- ``Carb Manager: Keto Diet Tracker \& Macros Counter", ``AI Nutrition Tracker: Macro Diet \& Calorie Counter", and ``TrackEats". 

We have also reviewed the ability to add new food items criterion for apps. As we have witnessed food consumption tracking and recommendation apps are not all enriched with food item databases, so, users cannot find the name of the food they have consumed in most of the apps. In such cases, some apps allow the user to add new food item names and the nutritional value to its database. In the manual process, 71.25\% of the apps (57/80) let users add food items. In the case of manually performing this task, some apps allow a new food item to be added to in the global database while others add food items to the local database. We found that some apps (e.g., ``Calorie Counter - MyFitnessPal", ``Calorie Counter, Carb Manager \& Keto by Freshbit") allow users to add new food names and nutritional information into their global database, which means other users can access that newly added food item information. On the other hand, some apps (e.g., ``Calorie Counter by FatSecret", ``Keto.app - Keto diet tracker") allow users to add new food items only into that apps' local database and, not in the main database. Therefore, if users add any new food items, only that specific user can see the added food items in the app. Users have to provide total nutritional values, and sometimes an image or barcode number to add the food into the app's database. The rest of the apps do not allow users to add foods to their database.

Table~\ref{feature} reports the percentage of functionality measurement criteria fulfilled by the apps evaluated from the three app stores. It provides an overall quality rating of all the apps for the different platforms (app stores). The results in Table~\ref{feature} show that the Google Play store (i.e., Android platform) contains a large number of apps compared to the Apple app and Microsoft stores. It also shows that the food recommendation criterion is rarely present among the 80 apps -- only 4 apps could recommend foods to users. 
\begin{table}
    \centering
    \caption{Assessment criteria for the measurement functionality of apps in three different app platforms}
    \begin{tabular}{ lcccc}
 \hline
 \label{feature}
 Measurement criteria & \makecell[c]{Google Play (n=70)\\n (\%)} & \makecell[c]{Microsoft (n=6)\\n (\%)} & \makecell[c]{Apple app (n=4)\\n (\%)} & \makecell[c]{Total (N=80)\\N (\%)}\\
 \hline\hline
Food recognition& 65 (92.86\%) &6 (100\%)&   3 (75\%)&74 (92.5\%)\\
Volume estimation&67 (95.71\%)&4 (66.67\%)&2 (50\%)&73 (91.25\%)\\
Nutrition estimation &64 (91.43\%)&5 (83.33\%)&2 (50\%)&71 (88.75\%)\\
History visualization &68 (97.14\%)&2 (33.33\%)&3 (75\%)&73 (91.25\%)\\
Food recommendation&3 (4.29\%)&0 (0\%)&1 (25\%)&4 (5\%)\\
Allow to add new food info&52 (74.29\%)&2 (33.33\%)&3 (75\%)&57 (71.25\%)\\
\hline

\hline
\end{tabular}
\end{table}

\subsection{Analysis of user reviews from the app stores}
User reviews or comments in the app stores play a vital role in identifying any app's quality~\citep{guzman2018user}. These reviews often provide detailed information about the features of the apps and their pros and cons. Thus, many people rely on users' reviews before downloading an app, as these reviews act as quality indicators of apps~\citep{vasa2012preliminary}. In the commercial app world, developers review the positive and negative comments to help improve their apps. 

\begin{figure}[!htbp]
\centering
  \begin{subfigure}[b]{.49\textwidth}
    \includegraphics[width=\textwidth]{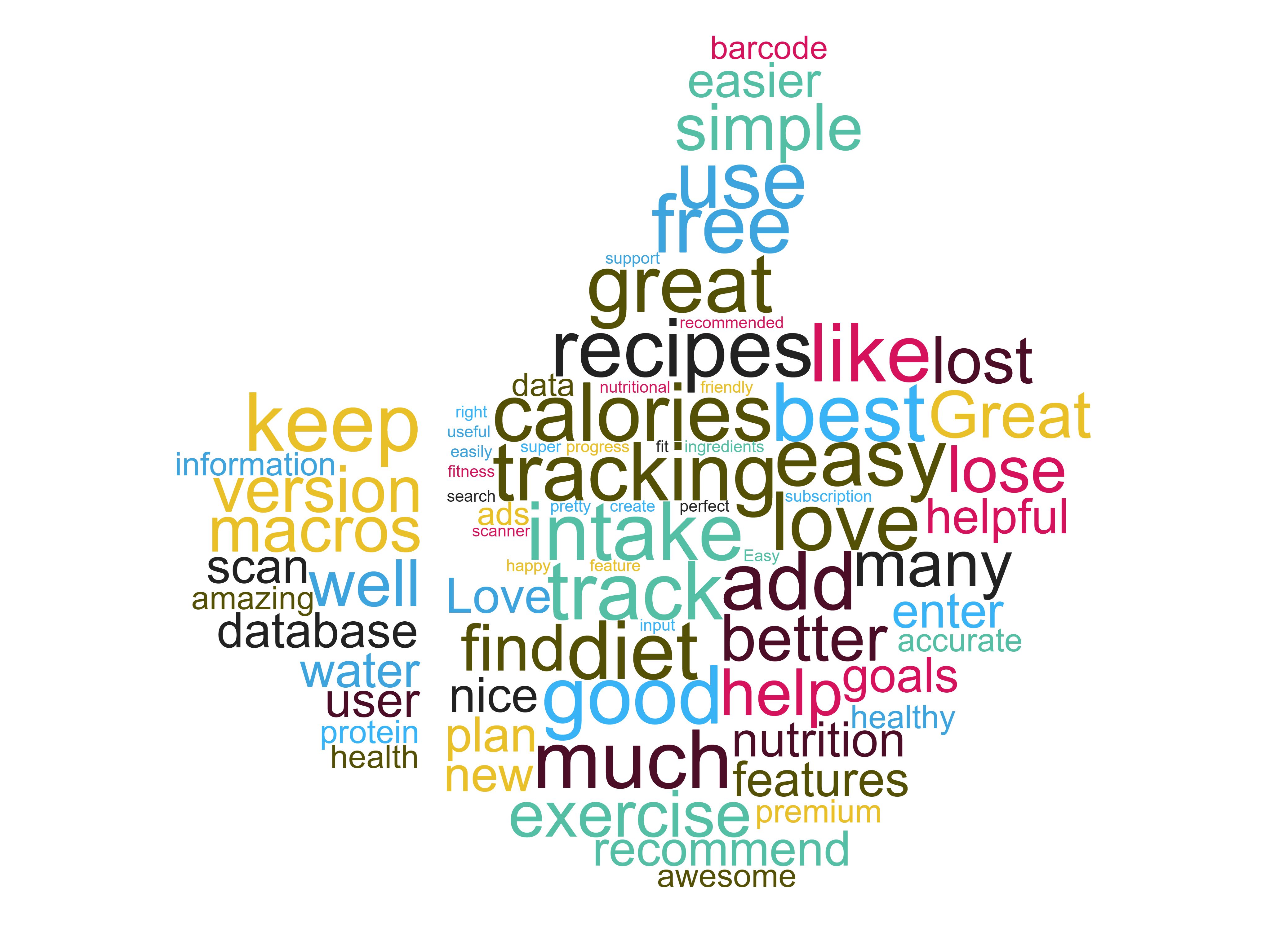}
    \caption{Positive comments}
    \label{good}
  \end{subfigure}
  \hfill
  \begin{subfigure}[b]{0.49\textwidth}
    \includegraphics[width=\textwidth]{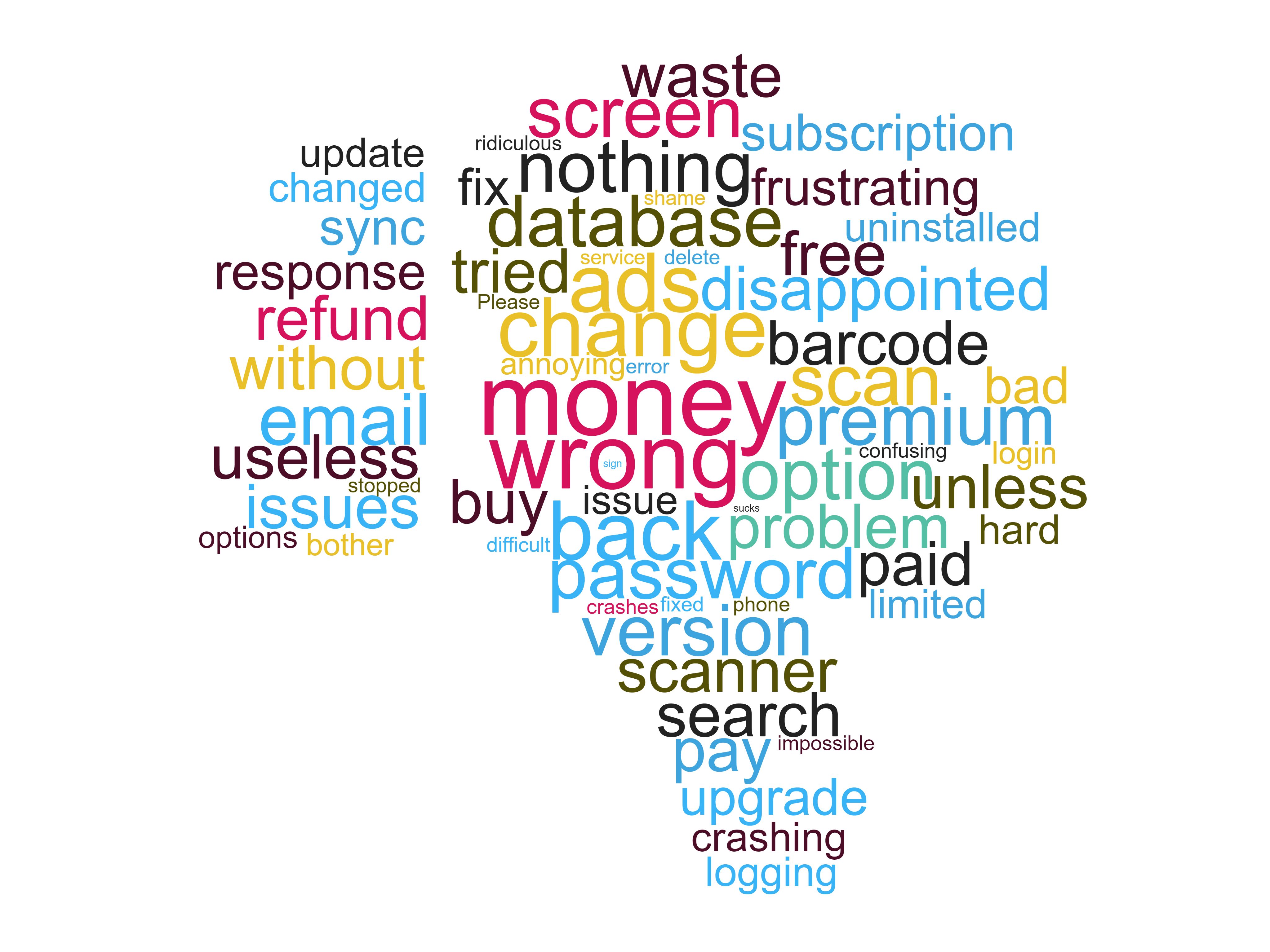}
    \caption{Negative comments}
    \label{bad}
  \end{subfigure}
  \caption{Word cloud of positive and negative comments}
  \label{Comment comparison}
\end{figure}

Considering the importance of user review analysis, in this study, we collected users' comments for the apps from their respective app stores. While collecting comments, we classified them into two categories based on the users' ratings for the apps: a comment is classified as ``positive" if the commenter rated the app four stars or above; otherwise the comment is classified as ``negative" review.
We have collected 900 user reviews, among which 55\% were positive and the rest were negative. Usually, in positive reviews, users tend to point out the features they liked most in the apps, and how benefited by using the apps. On the contrary, the negative reviews mainly reflect the limitations, faulty features or inaccurate descriptions of the apps.We used a word-cloud for visualizing comments to gain better insights from the users' reviews. Figure~\ref{Comment comparison} depicts the word-cloud of the positive and negative reviews.

Figure~\ref{good} depicts the word-cloud for positive reviews, highlighting some of the most frequently appearing positive words. The word ``like" is the most frequent in the positive reviews. This word is frequently used along with some other frequent words such as ``track", ``calories", ``exercise", and ``intake" because users like these components of the apps (i.e., weekly history visualization, track calorie intake, tracking exercise). Many users commented that they loved the apps for having an option to add food recipes into the database and recommend a diet plan. Therefore, we have identified the words ``love", ``add", ``recipes", ``database",``recommend", ``diet", and ``plan" as more recurring words. Users expressed their feelings using words such as ``great", ``good", and ``much" because some apps' designs are very simple and they were easy to use. As a result, we saw the words ``simple", ``easy", and ``use" in the good comments. Users like the apps that have a barcode scanner feature to estimate nutritional values like macros and, protein. Hence the words ``barcode", ``nutrition", ``macros", ``protein" are occur frequently in the positive reviews. Many apps provide some diet plans that can help users to lose weight. Thus, we see ``helpful" and ``lose" in the users' positive reviews. Some users said that they wanted to recommend the apps to their friends. Some users expressed in the good comments that the apps helped them to set goals and could track their health progress, including tracking their weight. Most of the users said that they liked the free version of most of the apps. For this reason we see the words ``free" and ``version" in our word-cloud of positive reviews.

Figure~\ref{bad} depicts the most frequently used words in the negative reviews. Here, we witnessed words such as ``money", ``paid", ``refund", ``subscription", ``waste" and ``premium". Users are not satisfied because components do not work properly in the paid apps or the apps having premium subscription packages. For these categories of apps, users felt that spending money on them apps was a total waste and they wanted refund. Frequent ads in the free apps disappoint the users and for this we see words such as ``free", ``ads", ``disappointed" in the word-cloud. In some apps, the barcode scanners do not work properly. Therefore, ``barcode", ``scan" have occurred in the users' negative reviews. Some users complained about problems they experienced while trying to login from another phone. A few users expressed confusion about the design and layout while they were using the apps. In some comments we noticed problems such as, users becoming frustrated by an app crashing, not letting users change their passwords, login issues when changing devices even though the email address is same, wrong estimations of calories or nutrition, etc. As a result, they uninstalled the apps. That is why, in the word-cloud we see some words like -- ``frustrating", ``useless", ``hard", ``crashing", ``change", ``bother", ``issue", ``password", ``email", ``wrong", ``uninstalled", etc.


\section{Findings and discussion}\label{discus}
\subsection{Limitations of reviewed apps}
We outline the major shortcomings of the reviewed apps for food consumption tracking and recommendation below:
\begin{enumerate}[label=(\roman*)]
\item \textbf{Lack of automatic food computation features:}
In our findings most of the apps are for the tracking user’s' consumption of foods. The apps need to recognize the food, estimate the food's portion size, and estimate nutritional values to track food consumption. However, in most of the apps, these tasks are done manually. Therefore, users must manually input these values to track their daily consumption. We find that only a few apps can recognize food directly from images, but these apps also have some issues. For instance, they first anticipate the ingredients of the food items, which is not always accurate. As a result, calorie and nutrient value estimations performed by the apps are inaccurate. In some cases, we noticed that if the food image of has multiple food items, then the apps focused only on the main item, and the rest of the items were ignored. For example, in a rice dish with side vegetables and, sauces, most apps only recognize the rice, as it is the main item, with a bigger portion. However, it is not enough to only recognize one food item among so many items. Moreover, in the case of low-quality images, this feature of the apps stops working. We have observed underestimation when it comes to computing food volumes. Most of the apps require manual entry for food volumes; only a few have the option to compute volume, which is often error-prone as well.

\item \textbf{Scarce existing database:}
 We spotted a deficiency in the existing food databases used in the commercial apps. We found two types of databases: international databas, that contain only international foods common in any country, for example, fruits, meat, fish, and vegetables, and the apps' country-specific food items. In the other database, only region-wise foods like Asian foods or European foods were listed. For this reason, when a user searches foods from the app’s databases manually or through barcode scanning, they cannot always find the foods. In the case of region-specific database apps, users from different regions cannot find their country-specific foods. Also, the apps that only focused on international food items, had a small list of country-specific food items in their databases, so the users could not find their cultural foods. Some of the apps give users an option to add foods. Even though by using this feature the users can add foods in the apps, the newly added foods are confined to that particular device or individual account. Other users cannot access that food item. We have found a few apps that allow users to add food items to their main databases. In these cases, users are asked to send several images of the food product and after a verification process, the apps' server adds the food item into the main database.
 
\item \textbf{Lack of evidence-based app:}
Evidence-based strategies are critical for food consumption tracking and recommendation apps. Food directly contributes to human health; thus, any information related to food needs to be verified, as wrong information can cause health risks or diseases. Hence, food consumption tracking and recommendation apps necessitate the involvement of dietitians, health experts, or nutritionists. This involvement raises the level of evidence or integrity of these apps. However, evidence-based strategies mostly were absent in the mobile apps we reviewed. Evidence-based apps are those that have been verified in a scientific peer-reviewed journal. There are only 12 apps that are based on evidence (e.g.,``Lose it!", ``My Fitness Pal", and ``Lifesum - Diet Plan, Macro Calculator \& Food Diary''). When an app suggests a diet plan to a user, the involvement of an expert or dietitian is a must but we rarely found this association in an app. We also searched for apps that can recommend food to users. We found four apps with this recommendation feature, (``Carb Manager: Keto Diet Tracker \& Macros Counter'', ``AI Nutrition Tracker: Macro Diet \& Calorie Counter'', ``AI Nutrition Tracker: Macro Diet \& Calorie Counter'', and ``WW Weight Watchers Reimagined'' ), but even in these apps the involvement of a specialist was missing. 
\end{enumerate}

\subsection{Design considerations}
 We have illustrated some future directions from app developers' and users' perspectives. We have also suggested some functionality improvements based on our findings.
\begin{enumerate}[label=(\roman*)]
    \item \textbf{Food computing automation improvement:} The recognition of food items, estimation of volume and nutrition from photos should exist in the apps and need to function properly. Various deep learning algorithms have been proposed to recognize food items from images~\citep{min2019survey}. There has been  extensive research in the field of food recognition~\citep{chopra2021recent,NAYAK2020100297,10.1145/3391624,mezgec2021deep,van2020use}. \cite{yang2010food} used the semantic texton forest to categorize all image pixels and then extracted the pairwise feature distribution as visual features. Nowadays, CNNs such as AlexNet~\citep{kagaya2014food}, GoogLeNet~\citep{wu2016learning}, Network-In-Networks (NIN)~\citep{tanno2016deepfoodcam}, Inception V3~\citep{hassannejad2016food}, ResNet~\citep{ming2018food}, and their combinations, are frequently employed for feature extraction in food identification~\citep{mcallister2018combining,pandey2017foodnet}. In a real-world scenario, there will be many food items present in one image.  There are some studies on multi-label ingredient recognition~\citep{bolanos2016simultaneous,chen2018fast}. \cite{aguilar2018grab} introduced a semantic food detection framework that consists of three parts: food segmentation~\citep{chopra2021recent}, food detection, and semantic food detection. Nevertheless, while reviewing the apps, we observed the absence of automatic food recognition, volume estimation, and food recommendation features in most of the apps. For food volume estimation~\citep{healthcare9121676}, one study proposed a three-stage system to calculate portion sizes using only two images of a dish acquired by mobile devices~\citep{dehais2016two}. Besides the CNNs, the generative adversarial networks are also used for food portion estimation~\citep{fang2018single}. \cite{kong2012dietcam} presented a mobile phone-based system, DietCam, which only requires users to take three images or a short video of the meal. Some research have focused on food recommendation systems~\citep{8930090}. During the  recommendations, users’ food preferences and health requirements should be considered~\citep{10.1145/3418211}. In~\cite{phanich2010food}, the authors introduced the Food Recommendation System (FRS) for diabetic patients, utilizing food clustering analysis. In the context of nutrition~\citep{KIRK2021104365} and food features, their system offered the appropriate substitute foods. In~\cite{jiang2019market2dish}, the authors revealed a personalized health-aware food recommendation system that could recognize the ingredients in market micro-videos, profile users' health status from their social media accounts, and offer personalized healthy foods to users. Hence, as we observed, implementing these automatic features in food consumption tracking and recommendation apps is essential yet challenging for developers. Therefore, developers should focus on these features to implement in future app development. 

\item \textbf{Use of enriched food database:}
 Food databases should be enriched to identify any food items. Many food databases are available, like -- ``Food-101" which is a public dataset consisting of 101,000 images with 101 categories~\citep{bossard2014food}, ``Food201 segmented dataset" which has 201 food classes consisting of 12,093 data items~\citep{meyers2015im2calories}, and ``Recipe1M+" which is a new structured database with over one million cooking recipes and 13 million food photos~\citep{marin2019recipe1m+}. However, we have seen that some of the apps' databases contain only international food items like fruit, vegetable, meat or fish. These items are common in all countries. But some apps were country of origin specific. Their database mostly had the food items of the apps' specific country. So, users from other countries find it difficult to use these apps, as there are fewer known food items in the apps' database. We barely found any apps with a highly enriched database with a vast number of food items. As we have seen, the databases used in the presented food consumption tracking and recommendation apps are not very rich. In the future, developers can utilize the datasets mentioned above to improve food consumption tracking and recommendation apps.
\item \textbf{Expert involvement:}
Dietitians are professionals in diet and nutrition who have had extensive training and experience. Also, they are well-known for providing successful lifestyle strategies for weight management via health behavior consulting~\citep{jensen20142013, millen20142013}. Moreover,the inclusion of health care experts in the development of medical urology apps has been found to have a favorable impact on app downloads, implying that working with health care experts provides users with greater confidence in the apps' safety and legitimacy~\citep{jospe2015diet}. Additionally, because dietitians use smartphone health applications and other mHealth technology in patient care~\citep{lieffers2014use,chen2017use,jospe2015diet}, their involvement in food consumption tracking and recommendations app is also expected. But unfortunately, the involvement of expert dietitians or nutritionists is insufficient in the existing apps, so the developers should consider increasing the involvement of medical experts in this app domain.

\item \textbf{Improvement of software qualities:}
Because the user interface connects the customer to the service they require, it is one of the most critical aspects to consider when designing and developing a commercial app~\citep{faghih2014user}. The user interface design determines whether a software application will stand or fall. Details of how to navigate the app and its services  must be user-friendly; otherwise, the user will be unable to traverse the program~\citep{ross2016overcoming}. That is why this feature needs to be considered in an app's design. Furthermore, various machine learning methods may be employed for food automation activities so that app performance (total battery life impact, chance of device heating) does not suffer. Transparency is a critical component of every mobile app. No app would be trusted unless it has sufficient credibility~\citep{corral2014defining}. Thus, the sub-scale of transparency in terms of user consent, the accuracy of the store description, the validity of the source, and the practicality of fulfilling the stated goals, must all be considered. In the case of free apps, advertisements are troublesome for users. Therefore, apps should be made ad-free or show fewer advertisements.

\end{enumerate} 

\subsection{Limitations of this study}
A limitation of this study, is that we only evaluated English language apps and did not consider apps that have access restrictions by region. Again, the three raters had three different mobile devices with different operating systems, so some apps worked differently on different devices. Their ratings were also different from each other for some criteria in our app rating scale. In our rating tool, there is a sub-scale named perceived impact on the user that consists of awareness induction behavior, knowledge enhancing behaviour, change of attitude toward improving balanced diet, intention to change, balanced diet related help-seeking behavior, and behaviour change of the users. As these criteria are mainly qualitative, their evaluation was subjective to the raters. Since our search and evaluation, some of the apps might have been removed from the app stores, or been updated with enhanced functionality. Also, new apps might have been  added to the app stores.

\section{ Conclusion}\label{conclusion}
In this study, we conducted a critical review of mobile apps from three popular app stores. Our search results identified a total of 473 related apps, from which we selected and evaluated 80 apps using our modified app rating tool. We devised this app rating tool specifically for analyzing food consumption tracking and recommendation apps by adopting and extending existing mobile app rating scales. Using this rating tool, we evaluated the selected 80 apps and analysed and identified their design faults. According to our evaluation, most of the existing mobile apps in the app stores do not meet the essential requirements for correctly tracking food consumption and recommendations. Although a few apps had some of the expected features, none met all the required functionalities. For most of the apps, tracking information required manual data input. The databases that are used in the apps are not enriched. We also observed that there are very few  evidence-based apps. Because there have been numerous studies about automatic food recognition, food portion size estimates, and nutritional value assessments, these aspects must be included in modern food consumption tracking and recommendation apps. Also, there has been much research on food recommendations but this feature is absent in most of the evaluated apps, that is why this feature needs to be included in future apps. These apps suggest diet plans, recommend foods to users, and estimate nutrient values, so an expert dietitian or nutritionist should be involved in their development. Also, enrichment of the database is required as nowadays multiple food datasets are available. Software qualities (aesthetics, general features, performance, usability)  also play a vital role in commercial apps and thus developers need to consider these matters.

Nonetheless, the analysis provided here covers a variety of general quality features and specific functional features that can be used in food consumption tracking and recommendation apps to provide consumers with a realistic and evidence-based experience. Studies show how people use smartphones to improve their fitness and obesity literacy, as well as the overall status of the commercial product market for food consumption tracking and recommendation apps. This study will open the door to future researchers who focus on the implementation, effectiveness and performance measurement of food computing apps.

\section*{CRediT authorship contribution statement}
\textbf{S. Samad, F. Ahmed, S. Naher}: Conceptualization, Data curation, Formal analysis, Writing -- original draft. \textbf{M. A. Kabir:} Conceptualization, Data curation, Methodology, Supervision, Visualization, Writing -- review \& editing. \textbf{A. Das and S. Amin:} Data curation, Writing -- review \& editing. \textbf{S. M. S. Islam:} Writing -- review \& editing.

\section*{Declaration of Competing Interest}
The authors declare that they have no known competing financial interests or personal relationships that could have appeared to influence the work reported in this paper.

\section*{Funding Sources}
This research did not receive any specific grant from funding agencies in the public, commercial, or not-for-profit sectors.

\section*{Acknowledgment}
The authors would like to thank Golam Sarwar Tuhin (CUET) for his help with the app ratings. 



\bibliographystyle{model5-names}
\biboptions{authoryear}
\bibliography{mybib}

\begin{thebibliography}{106}
\expandafter\ifx\csname natexlab\endcsname\relax\def\natexlab#1{#1}\fi
\providecommand{\url}[1]{\texttt{#1}}
\providecommand{\href}[2]{#2}
\providecommand{\path}[1]{#1}
\providecommand{\DOIprefix}{doi:}
\providecommand{\ArXivprefix}{arXiv:}
\providecommand{\URLprefix}{URL: }
\providecommand{\Pubmedprefix}{pmid:}
\providecommand{\doi}[1]{\href{http://dx.doi.org/#1}{\path{#1}}}
\providecommand{\Pubmed}[1]{\href{pmid:#1}{\path{#1}}}
\providecommand{\bibinfo}[2]{#2}
\ifx\xfnm\relax \def\xfnm[#1]{\unskip,\space#1}\fi
\bibitem[{Aguilar et~al.(2019)Aguilar, Bola{\~n}os \&
  Radeva}]{aguilar2019regularized}
\bibinfo{author}{Aguilar, E.}, \bibinfo{author}{Bola{\~n}os, M.}, \&
  \bibinfo{author}{Radeva, P.} (\bibinfo{year}{2019}).
\newblock \bibinfo{title}{Regularized uncertainty-based multi-task learning
  model for food analysis}.
\newblock {\it \bibinfo{journal}{Journal of Visual Communication and Image
  Representation}\/},  {\it \bibinfo{volume}{60}\/}, \bibinfo{pages}{360--370}.
\bibitem[{Aguilar et~al.(2018)Aguilar, Remeseiro, Bola{\~n}os \&
  Radeva}]{aguilar2018grab}
\bibinfo{author}{Aguilar, E.}, \bibinfo{author}{Remeseiro, B.},
  \bibinfo{author}{Bola{\~n}os, M.}, \& \bibinfo{author}{Radeva, P.}
  (\bibinfo{year}{2018}).
\newblock \bibinfo{title}{Grab, pay, and eat: Semantic food detection for smart
  restaurants}.
\newblock {\it \bibinfo{journal}{IEEE Transactions on Multimedia}\/},  {\it
  \bibinfo{volume}{20}\/}, \bibinfo{pages}{3266--3275}.
\bibitem[{Anthimopoulos et~al.(2014)Anthimopoulos, Gianola, Scarnato, Diem \&
  Mougiakakou}]{anthimopoulos2014food}
\bibinfo{author}{Anthimopoulos, M.~M.}, \bibinfo{author}{Gianola, L.},
  \bibinfo{author}{Scarnato, L.}, \bibinfo{author}{Diem, P.}, \&
  \bibinfo{author}{Mougiakakou, S.~G.} (\bibinfo{year}{2014}).
\newblock \bibinfo{title}{A food recognition system for diabetic patients based
  on an optimized bag-of-features model}.
\newblock {\it \bibinfo{journal}{IEEE journal of biomedical and health
  informatics}\/},  {\it \bibinfo{volume}{18}\/}, \bibinfo{pages}{1261--1271}.
\bibitem[{Beal et~al.(2013)Beal, Decker \& Quelly}]{beal2013should}
\bibinfo{author}{Beal, J.}, \bibinfo{author}{Decker, J.~W.}, \&
  \bibinfo{author}{Quelly, S.} (\bibinfo{year}{2013}).
\newblock \bibinfo{title}{Should extreme obesity in children be considered
  child abuse?}
\newblock {\it \bibinfo{journal}{MCN: The American Journal of Maternal/Child
  Nursing}\/},  {\it \bibinfo{volume}{38}\/}, \bibinfo{pages}{334--335}.
\bibitem[{Boland \& Bronlund(2019)}]{BOLAND2019634}
\bibinfo{author}{Boland, M.}, \& \bibinfo{author}{Bronlund, J.}
  (\bibinfo{year}{2019}).
\newblock \bibinfo{title}{enutrition - the next dimension for ehealth?}
\newblock {\it \bibinfo{journal}{Trends in Food Science \& Technology}\/},
  {\it \bibinfo{volume}{91}\/}, \bibinfo{pages}{634--639}.
  \DOIprefix\doi{10.1016/j.tifs.2019.08.001}.
\bibitem[{Bolanos \& Radeva(2016)}]{bolanos2016simultaneous}
\bibinfo{author}{Bolanos, M.}, \& \bibinfo{author}{Radeva, P.}
  (\bibinfo{year}{2016}).
\newblock \bibinfo{title}{Simultaneous food localization and recognition}.
\newblock In {\it \bibinfo{booktitle}{2016 23rd International Conference on
  Pattern Recognition (ICPR)}\/} (pp. \bibinfo{pages}{3140--3145}).
\newblock \bibinfo{organization}{IEEE}.
\bibitem[{Bossard et~al.(2014)Bossard, Guillaumin \&
  Van~Gool}]{bossard2014food}
\bibinfo{author}{Bossard, L.}, \bibinfo{author}{Guillaumin, M.}, \&
  \bibinfo{author}{Van~Gool, L.} (\bibinfo{year}{2014}).
\newblock \bibinfo{title}{Food-101--mining discriminative components with
  random forests}.
\newblock In {\it \bibinfo{booktitle}{European conference on computer
  vision}\/} (pp. \bibinfo{pages}{446--461}).
\newblock \bibinfo{organization}{Springer}.
\bibitem[{Brug et~al.(1995)Brug, Debie, van Assema \&
  Weijts}]{brug1995psychosocial}
\bibinfo{author}{Brug, J.}, \bibinfo{author}{Debie, S.}, \bibinfo{author}{van
  Assema, P.}, \& \bibinfo{author}{Weijts, W.} (\bibinfo{year}{1995}).
\newblock \bibinfo{title}{Psychosocial determinants of fruit and vegetable
  consumption among adults: results of focus group interviews}.
\newblock {\it \bibinfo{journal}{Food Quality and preference}\/},  {\it
  \bibinfo{volume}{6}\/}, \bibinfo{pages}{99--107}.
\bibitem[{Chandra(1997)}]{chandra1997nutrition}
\bibinfo{author}{Chandra, R.~K.} (\bibinfo{year}{1997}).
\newblock \bibinfo{title}{Nutrition and the immune system: an introduction}.
\newblock {\it \bibinfo{journal}{The American journal of clinical
  nutrition}\/},  {\it \bibinfo{volume}{66}\/}, \bibinfo{pages}{460S--463S}.
\bibitem[{Chen et~al.(2018)Chen, Xu, Xiao, Wu \& Zhang}]{chen2018fast}
\bibinfo{author}{Chen, H.}, \bibinfo{author}{Xu, J.}, \bibinfo{author}{Xiao,
  G.}, \bibinfo{author}{Wu, Q.}, \& \bibinfo{author}{Zhang, S.}
  (\bibinfo{year}{2018}).
\newblock \bibinfo{title}{Fast auto-clean cnn model for online prediction of
  food materials}.
\newblock {\it \bibinfo{journal}{Journal of Parallel and Distributed
  Computing}\/},  {\it \bibinfo{volume}{117}\/}, \bibinfo{pages}{218--227}.
\bibitem[{Chen et~al.(2015)Chen, Cade \& Allman-Farinelli}]{chen2015most}
\bibinfo{author}{Chen, J.}, \bibinfo{author}{Cade, J.~E.}, \&
  \bibinfo{author}{Allman-Farinelli, M.} (\bibinfo{year}{2015}).
\newblock \bibinfo{title}{The most popular smartphone apps for weight loss: a
  quality assessment}.
\newblock {\it \bibinfo{journal}{JMIR mHealth and uHealth}\/},  {\it
  \bibinfo{volume}{3}\/}, \bibinfo{pages}{e104}.
\bibitem[{Chen et~al.(2017)Chen, Lieffers, Bauman, Hanning \&
  Allman-Farinelli}]{chen2017use}
\bibinfo{author}{Chen, J.}, \bibinfo{author}{Lieffers, J.},
  \bibinfo{author}{Bauman, A.}, \bibinfo{author}{Hanning, R.}, \&
  \bibinfo{author}{Allman-Farinelli, M.} (\bibinfo{year}{2017}).
\newblock \bibinfo{title}{The use of smartphone health apps and other mobile h
  ealth (mhealth) technologies in dietetic practice: a three country study}.
\newblock {\it \bibinfo{journal}{Journal of Human Nutrition and Dietetics}\/},
  {\it \bibinfo{volume}{30}\/}, \bibinfo{pages}{439--452}.
\bibitem[{Chetrari(2017)}]{chetrari2017characteristics}
\bibinfo{author}{Chetrari, A.} (\bibinfo{year}{2017}).
\newblock {\it \bibinfo{title}{Characteristics of value-providing consumer
  smartphone apps}\/}.
\newblock Ph.D. thesis Empire State College.
\bibitem[{Chopra \& Purwar(2021)}]{chopra2021recent}
\bibinfo{author}{Chopra, M.}, \& \bibinfo{author}{Purwar, A.}
  (\bibinfo{year}{2021}).
\newblock \bibinfo{title}{Recent studies on segmentation techniques for food
  recognition: A survey}.
\newblock {\it \bibinfo{journal}{Archives of Computational Methods in
  Engineering}\/},  (pp. \bibinfo{pages}{1--14}).
\bibitem[{Christmann \& Van~Aelst(2006)}]{christmann2006robust}
\bibinfo{author}{Christmann, A.}, \& \bibinfo{author}{Van~Aelst, S.}
  (\bibinfo{year}{2006}).
\newblock \bibinfo{title}{Robust estimation of cronbach's alpha}.
\newblock {\it \bibinfo{journal}{Journal of Multivariate Analysis}\/},  {\it
  \bibinfo{volume}{97}\/}, \bibinfo{pages}{1660--1674}.
\bibitem[{Christodoulidis et~al.(2015)Christodoulidis, Anthimopoulos \&
  Mougiakakou}]{christodoulidis2015food}
\bibinfo{author}{Christodoulidis, S.}, \bibinfo{author}{Anthimopoulos, M.}, \&
  \bibinfo{author}{Mougiakakou, S.} (\bibinfo{year}{2015}).
\newblock \bibinfo{title}{Food recognition for dietary assessment using deep
  convolutional neural networks}.
\newblock In {\it \bibinfo{booktitle}{International Conference on Image
  Analysis and Processing}\/} (pp. \bibinfo{pages}{458--465}).
\newblock \bibinfo{organization}{Springer}.
\bibitem[{Chu et~al.(2018)Chu, Nguyet, Dinh, Lien, Nguyen, Ngoc, Tao, Le, Nga,
  Jurgo{\'n}ski et~al.}]{chu2018update}
\bibinfo{author}{Chu, D.-T.}, \bibinfo{author}{Nguyet, N. T.~M.},
  \bibinfo{author}{Dinh, T.~C.}, \bibinfo{author}{Lien, N. V.~T.},
  \bibinfo{author}{Nguyen, K.-H.}, \bibinfo{author}{Ngoc, V. T.~N.},
  \bibinfo{author}{Tao, Y.}, \bibinfo{author}{Le, D.-H.}, \bibinfo{author}{Nga,
  V.~B.}, \bibinfo{author}{Jurgo{\'n}ski, A.} et~al. (\bibinfo{year}{2018}).
\newblock \bibinfo{title}{An update on physical health and economic
  consequences of overweight and obesity}.
\newblock {\it \bibinfo{journal}{Diabetes \& Metabolic Syndrome: Clinical
  Research \& Reviews}\/},  {\it \bibinfo{volume}{12}\/},
  \bibinfo{pages}{1095--1100}.
\bibitem[{Corral et~al.(2014)Corral, Sillitti \& Succi}]{corral2014defining}
\bibinfo{author}{Corral, L.}, \bibinfo{author}{Sillitti, A.}, \&
  \bibinfo{author}{Succi, G.} (\bibinfo{year}{2014}).
\newblock \bibinfo{title}{Defining relevant software quality characteristics
  from publishing policies of mobile app stores}.
\newblock In {\it \bibinfo{booktitle}{International Conference on Mobile Web
  and Information Systems}\/} (pp. \bibinfo{pages}{205--217}).
\newblock \bibinfo{organization}{Springer}.
\bibitem[{Cronbach(1951)}]{cronbach1951coefficient}
\bibinfo{author}{Cronbach, L.~J.} (\bibinfo{year}{1951}).
\newblock \bibinfo{title}{Coefficient alpha and the internal structure of
  tests}.
\newblock {\it \bibinfo{journal}{psychometrika}\/},  {\it
  \bibinfo{volume}{16}\/}, \bibinfo{pages}{297--334}.
\bibitem[{Darby et~al.(2016)Darby, Strum, Holmes \& Gatwood}]{darby2016review}
\bibinfo{author}{Darby, A.}, \bibinfo{author}{Strum, M.~W.},
  \bibinfo{author}{Holmes, E.}, \& \bibinfo{author}{Gatwood, J.}
  (\bibinfo{year}{2016}).
\newblock \bibinfo{title}{A review of nutritional tracking mobile applications
  for diabetes patient use}.
\newblock {\it \bibinfo{journal}{Diabetes technology \& therapeutics}\/},  {\it
  \bibinfo{volume}{18}\/}, \bibinfo{pages}{200--212}.
\bibitem[{Dehais et~al.(2016)Dehais, Anthimopoulos, Shevchik \&
  Mougiakakou}]{dehais2016two}
\bibinfo{author}{Dehais, J.}, \bibinfo{author}{Anthimopoulos, M.},
  \bibinfo{author}{Shevchik, S.}, \& \bibinfo{author}{Mougiakakou, S.}
  (\bibinfo{year}{2016}).
\newblock \bibinfo{title}{Two-view 3d reconstruction for food volume
  estimation}.
\newblock {\it \bibinfo{journal}{IEEE transactions on multimedia}\/},  {\it
  \bibinfo{volume}{19}\/}, \bibinfo{pages}{1090--1099}.
\bibitem[{Elsweiler et~al.(2015)Elsweiler, Harvey, Ludwig \&
  Said}]{elsweiler2015bringing}
\bibinfo{author}{Elsweiler, D.}, \bibinfo{author}{Harvey, M.},
  \bibinfo{author}{Ludwig, B.}, \& \bibinfo{author}{Said, A.}
  (\bibinfo{year}{2015}).
\newblock \bibinfo{title}{Bringing the" healthy" into food recommenders.}
\newblock In {\it \bibinfo{booktitle}{DMRS}\/} (pp. \bibinfo{pages}{33--36}).
\bibitem[{Faghih et~al.(2014)Faghih, Azadehfar, Reza, Katebi
  et~al.}]{faghih2014user}
\bibinfo{author}{Faghih, B.}, \bibinfo{author}{Azadehfar, D.},
  \bibinfo{author}{Reza, M.}, \bibinfo{author}{Katebi, P.} et~al.
  (\bibinfo{year}{2014}).
\newblock \bibinfo{title}{User interface design for e-learning software}.
\newblock {\it \bibinfo{journal}{arXiv preprint arXiv:1401.6365}\/}, .
\bibitem[{Fang et~al.(2018)Fang, Shao, Mao, Fu, Delp, Zhu, Kerr \&
  Boushey}]{fang2018single}
\bibinfo{author}{Fang, S.}, \bibinfo{author}{Shao, Z.}, \bibinfo{author}{Mao,
  R.}, \bibinfo{author}{Fu, C.}, \bibinfo{author}{Delp, E.~J.},
  \bibinfo{author}{Zhu, F.}, \bibinfo{author}{Kerr, D.~A.}, \&
  \bibinfo{author}{Boushey, C.~J.} (\bibinfo{year}{2018}).
\newblock \bibinfo{title}{Single-view food portion estimation: Learning
  image-to-energy mappings using generative adversarial networks}.
\newblock In {\it \bibinfo{booktitle}{2018 25th IEEE International Conference
  on Image Processing (ICIP)}\/} (pp. \bibinfo{pages}{251--255}).
\newblock \bibinfo{organization}{IEEE}.
\bibitem[{Farinella et~al.(2014)Farinella, Moltisanti \&
  Battiato}]{farinella2014classifying}
\bibinfo{author}{Farinella, G.~M.}, \bibinfo{author}{Moltisanti, M.}, \&
  \bibinfo{author}{Battiato, S.} (\bibinfo{year}{2014}).
\newblock \bibinfo{title}{Classifying food images represented as bag of
  textons}.
\newblock In {\it \bibinfo{booktitle}{2014 IEEE International Conference on
  Image Processing (ICIP)}\/} (pp. \bibinfo{pages}{5212--5216}).
\newblock \bibinfo{organization}{IEEE}.
\bibitem[{Ferrara et~al.(2019)Ferrara, Kim, Lin, Hua \&
  Seto}]{ferrara2019focused}
\bibinfo{author}{Ferrara, G.}, \bibinfo{author}{Kim, J.}, \bibinfo{author}{Lin,
  S.}, \bibinfo{author}{Hua, J.}, \& \bibinfo{author}{Seto, E.}
  (\bibinfo{year}{2019}).
\newblock \bibinfo{title}{A focused review of smartphone diet-tracking apps:
  usability, functionality, coherence with behavior change theory, and
  comparative validity of nutrient intake and energy estimates}.
\newblock {\it \bibinfo{journal}{JMIR mHealth and uHealth}\/},  {\it
  \bibinfo{volume}{7}\/}, \bibinfo{pages}{e9232}.
\bibitem[{Franco et~al.(2016)Franco, Fallaize, Lovegrove \&
  Hwang}]{franco2016popular}
\bibinfo{author}{Franco, R.~Z.}, \bibinfo{author}{Fallaize, R.},
  \bibinfo{author}{Lovegrove, J.~A.}, \& \bibinfo{author}{Hwang, F.}
  (\bibinfo{year}{2016}).
\newblock \bibinfo{title}{Popular nutrition-related mobile apps: a feature
  assessment}.
\newblock {\it \bibinfo{journal}{JMIR mHealth and uHealth}\/},  {\it
  \bibinfo{volume}{4}\/}, \bibinfo{pages}{e85}.
\bibitem[{Friesen et~al.(2013)Friesen, Hamel \& McLeod}]{friesen2013mhealth}
\bibinfo{author}{Friesen, M.~R.}, \bibinfo{author}{Hamel, C.}, \&
  \bibinfo{author}{McLeod, R.~D.} (\bibinfo{year}{2013}).
\newblock \bibinfo{title}{A mhealth application for chronic wound care:
  Findings of a user trial}.
\newblock {\it \bibinfo{journal}{International journal of environmental
  research and public health}\/},  {\it \bibinfo{volume}{10}\/},
  \bibinfo{pages}{6199--6214}.
\bibitem[{Ge et~al.(2015)Ge, Ricci \& Massimo}]{ge2015health}
\bibinfo{author}{Ge, M.}, \bibinfo{author}{Ricci, F.}, \&
  \bibinfo{author}{Massimo, D.} (\bibinfo{year}{2015}).
\newblock \bibinfo{title}{Health-aware food recommender system}.
\newblock In {\it \bibinfo{booktitle}{Proceedings of the 9th ACM Conference on
  Recommender Systems}\/} (pp. \bibinfo{pages}{333--334}).
\bibitem[{Georgieva et~al.(2011)Georgieva, Smrikarov \&
  Georgiev}]{georgieva2011evaluation}
\bibinfo{author}{Georgieva, E.~S.}, \bibinfo{author}{Smrikarov, A.~S.}, \&
  \bibinfo{author}{Georgiev, T.~S.} (\bibinfo{year}{2011}).
\newblock \bibinfo{title}{Evaluation of mobile learning system}.
\newblock {\it \bibinfo{journal}{Procedia Computer Science}\/},  {\it
  \bibinfo{volume}{3}\/}, \bibinfo{pages}{632--637}.
\bibitem[{Gliem \& Gliem(2003)}]{gliem2003calculating}
\bibinfo{author}{Gliem, J.~A.}, \& \bibinfo{author}{Gliem, R.~R.}
  (\bibinfo{year}{2003}).
\newblock \bibinfo{title}{Calculating, interpreting, and reporting cronbach’s
  alpha reliability coefficient for likert-type scales}.
\newblock \bibinfo{organization}{Midwest Research-to-Practice Conference in
  Adult, Continuing, and Community~…}.
\bibitem[{Guzman et~al.(2018)Guzman, Oliveira, Steiner, Wagner \&
  Glinz}]{guzman2018user}
\bibinfo{author}{Guzman, E.}, \bibinfo{author}{Oliveira, L.},
  \bibinfo{author}{Steiner, Y.}, \bibinfo{author}{Wagner, L.~C.}, \&
  \bibinfo{author}{Glinz, M.} (\bibinfo{year}{2018}).
\newblock \bibinfo{title}{User feedback in the app store: a cross-cultural
  study}.
\newblock In {\it \bibinfo{booktitle}{2018 IEEE/ACM 40th International
  Conference on Software Engineering: Software Engineering in Society
  (ICSE-SEIS)}\/} (pp. \bibinfo{pages}{13--22}).
\newblock \bibinfo{organization}{IEEE}.
\bibitem[{Hassannejad et~al.(2016)Hassannejad, Matrella, Ciampolini, De~Munari,
  Mordonini \& Cagnoni}]{hassannejad2016food}
\bibinfo{author}{Hassannejad, H.}, \bibinfo{author}{Matrella, G.},
  \bibinfo{author}{Ciampolini, P.}, \bibinfo{author}{De~Munari, I.},
  \bibinfo{author}{Mordonini, M.}, \& \bibinfo{author}{Cagnoni, S.}
  (\bibinfo{year}{2016}).
\newblock \bibinfo{title}{Food image recognition using very deep convolutional
  networks}.
\newblock In {\it \bibinfo{booktitle}{Proceedings of the 2nd International
  Workshop on Multimedia Assisted Dietary Management}\/} (pp.
  \bibinfo{pages}{41--49}).
\bibitem[{He et~al.(2015)He, Kong \& Tan}]{he2015dietcam}
\bibinfo{author}{He, H.}, \bibinfo{author}{Kong, F.}, \& \bibinfo{author}{Tan,
  J.} (\bibinfo{year}{2015}).
\newblock \bibinfo{title}{Dietcam: multiview food recognition using a
  multikernel svm}.
\newblock {\it \bibinfo{journal}{IEEE journal of biomedical and health
  informatics}\/},  {\it \bibinfo{volume}{20}\/}, \bibinfo{pages}{848--855}.
\bibitem[{He et~al.(2014)He, Xu, Khanna, Boushey \& Delp}]{he2014analysis}
\bibinfo{author}{He, Y.}, \bibinfo{author}{Xu, C.}, \bibinfo{author}{Khanna,
  N.}, \bibinfo{author}{Boushey, C.~J.}, \& \bibinfo{author}{Delp, E.~J.}
  (\bibinfo{year}{2014}).
\newblock \bibinfo{title}{Analysis of food images: Features and
  classification}.
\newblock In {\it \bibinfo{booktitle}{2014 IEEE International Conference on
  Image Processing (ICIP)}\/} (pp. \bibinfo{pages}{2744--2748}).
\newblock \bibinfo{organization}{IEEE}.
\bibitem[{Hingle et~al.(2013)Hingle, Yoon, Fowler, Kobourov, Schneider, Falk \&
  Burd}]{hingle2013collection}
\bibinfo{author}{Hingle, M.}, \bibinfo{author}{Yoon, D.},
  \bibinfo{author}{Fowler, J.}, \bibinfo{author}{Kobourov, S.},
  \bibinfo{author}{Schneider, M.~L.}, \bibinfo{author}{Falk, D.}, \&
  \bibinfo{author}{Burd, R.} (\bibinfo{year}{2013}).
\newblock \bibinfo{title}{Collection and visualization of dietary behavior and
  reasons for eating using twitter}.
\newblock {\it \bibinfo{journal}{Journal of medical Internet research}\/},
  {\it \bibinfo{volume}{15}\/}, \bibinfo{pages}{e125}.
\bibitem[{Huebner et~al.(2019)Huebner, Schmid, Bouguerra \&
  Ilic}]{huebner2019finmars}
\bibinfo{author}{Huebner, J.}, \bibinfo{author}{Schmid, C.},
  \bibinfo{author}{Bouguerra, M.}, \& \bibinfo{author}{Ilic, A.}
  (\bibinfo{year}{2019}).
\newblock \bibinfo{title}{Finmars: A mobile app rating scale for finance apps}.
\newblock In {\it \bibinfo{booktitle}{Proceedings of the 9th International
  Conference on Information Communication and Management}\/} (pp.
  \bibinfo{pages}{6--11}).
\bibitem[{Hussain et~al.(2017)Hussain, Mkpojiogu, Musa \&
  Mortada}]{hussain2017user}
\bibinfo{author}{Hussain, A.}, \bibinfo{author}{Mkpojiogu, E.~O.},
  \bibinfo{author}{Musa, J.}, \& \bibinfo{author}{Mortada, S.}
  (\bibinfo{year}{2017}).
\newblock \bibinfo{title}{A user experience evaluation of amazon kindle mobile
  application}.
\newblock In {\it \bibinfo{booktitle}{AIP conference proceedings}\/} (p.
  \bibinfo{pages}{020060}).
\newblock \bibinfo{organization}{AIP Publishing LLC} volume
  \bibinfo{volume}{1891}.
\bibitem[{Jensen et~al.(2014)Jensen, Ryan, Apovian, Ard, Comuzzie, Donato, Hu,
  Hubbard, Jakicic, Kushner et~al.}]{jensen20142013}
\bibinfo{author}{Jensen, M.~D.}, \bibinfo{author}{Ryan, D.~H.},
  \bibinfo{author}{Apovian, C.~M.}, \bibinfo{author}{Ard, J.~D.},
  \bibinfo{author}{Comuzzie, A.~G.}, \bibinfo{author}{Donato, K.~A.},
  \bibinfo{author}{Hu, F.~B.}, \bibinfo{author}{Hubbard, V.~S.},
  \bibinfo{author}{Jakicic, J.~M.}, \bibinfo{author}{Kushner, R.~F.} et~al.
  (\bibinfo{year}{2014}).
\newblock \bibinfo{title}{2013 aha/acc/tos guideline for the management of
  overweight and obesity in adults: a report of the american college of
  cardiology/american heart association task force on practice guidelines and
  the obesity society}.
\newblock {\it \bibinfo{journal}{Journal of the American college of
  cardiology}\/},  {\it \bibinfo{volume}{63}\/}, \bibinfo{pages}{2985--3023}.
\bibitem[{Jiang et~al.(2019)Jiang, Wang, Liu, Nie, Duan \&
  Xu}]{jiang2019market2dish}
\bibinfo{author}{Jiang, H.}, \bibinfo{author}{Wang, W.}, \bibinfo{author}{Liu,
  M.}, \bibinfo{author}{Nie, L.}, \bibinfo{author}{Duan, L.-Y.}, \&
  \bibinfo{author}{Xu, C.} (\bibinfo{year}{2019}).
\newblock \bibinfo{title}{Market2dish: A health-aware food recommendation
  system}.
\newblock In {\it \bibinfo{booktitle}{Proceedings of the 27th ACM International
  Conference on Multimedia}\/} (pp. \bibinfo{pages}{2188--2190}).
\bibitem[{Jiang et~al.(2020)Jiang, Min, Lyu \& Liu}]{10.1145/3391624}
\bibinfo{author}{Jiang, S.}, \bibinfo{author}{Min, W.}, \bibinfo{author}{Lyu,
  Y.}, \& \bibinfo{author}{Liu, L.} (\bibinfo{year}{2020}).
\newblock \bibinfo{title}{Few-shot food recognition via multi-view
  representation learning}.
\newblock {\it \bibinfo{journal}{ACM Trans. Multimedia Comput. Commun.
  Appl.}\/},  {\it \bibinfo{volume}{16}\/}. \DOIprefix\doi{10.1145/3391624}.
\bibitem[{Jospe et~al.(2015)Jospe, Fairbairn, Green \& Perry}]{jospe2015diet}
\bibinfo{author}{Jospe, M.~R.}, \bibinfo{author}{Fairbairn, K.~A.},
  \bibinfo{author}{Green, P.}, \& \bibinfo{author}{Perry, T.~L.}
  (\bibinfo{year}{2015}).
\newblock \bibinfo{title}{Diet app use by sports dietitians: a survey in five
  countries}.
\newblock {\it \bibinfo{journal}{JMIR mHealth and uHealth}\/},  {\it
  \bibinfo{volume}{3}\/}, \bibinfo{pages}{e3345}.
\bibitem[{Kabir et~al.(2021)Kabir, Rahman, Islam, Ahmed \&
  Laird}]{ashad2020mobile}
\bibinfo{author}{Kabir, M.~A.}, \bibinfo{author}{Rahman, S.},
  \bibinfo{author}{Islam, M.~M.}, \bibinfo{author}{Ahmed, S.}, \&
  \bibinfo{author}{Laird, C.} (\bibinfo{year}{2021}).
\newblock \bibinfo{title}{Mobile apps for foot measurement in pedorthic
  practice: Scoping review}.
\newblock {\it \bibinfo{journal}{JMIR Mhealth Uhealth}\/},  {\it
  \bibinfo{volume}{9}\/}.
\bibitem[{Kagaya et~al.(2014)Kagaya, Aizawa \& Ogawa}]{kagaya2014food}
\bibinfo{author}{Kagaya, H.}, \bibinfo{author}{Aizawa, K.}, \&
  \bibinfo{author}{Ogawa, M.} (\bibinfo{year}{2014}).
\newblock \bibinfo{title}{Food detection and recognition using convolutional
  neural network}.
\newblock In {\it \bibinfo{booktitle}{Proceedings of the 22nd ACM international
  conference on Multimedia}\/} (pp. \bibinfo{pages}{1085--1088}).
\bibitem[{Kalinowska et~al.(2021)Kalinowska, Wojnowski \&
  Tobiszewski}]{KALINOWSKA2021271}
\bibinfo{author}{Kalinowska, K.}, \bibinfo{author}{Wojnowski, W.}, \&
  \bibinfo{author}{Tobiszewski, M.} (\bibinfo{year}{2021}).
\newblock \bibinfo{title}{Smartphones as tools for equitable food quality
  assessment}.
\newblock {\it \bibinfo{journal}{Trends in Food Science \& Technology}\/},
  {\it \bibinfo{volume}{111}\/}, \bibinfo{pages}{271--279}.
  \DOIprefix\doi{10.1016/j.tifs.2021.02.068}.
\bibitem[{Kallio et~al.(2005)Kallio, Kaikkonen et~al.}]{kallio2005usability}
\bibinfo{author}{Kallio, T.}, \bibinfo{author}{Kaikkonen, A.} et~al.
  (\bibinfo{year}{2005}).
\newblock \bibinfo{title}{Usability testing of mobile applications: A
  comparison between laboratory and field testing}.
\newblock {\it \bibinfo{journal}{Journal of Usability studies}\/},  {\it
  \bibinfo{volume}{1}\/}, \bibinfo{pages}{23--28}.
\bibitem[{Kawano \& Yanai(2014)}]{kawano2014food}
\bibinfo{author}{Kawano, Y.}, \& \bibinfo{author}{Yanai, K.}
  (\bibinfo{year}{2014}).
\newblock \bibinfo{title}{Food image recognition with deep convolutional
  features}.
\newblock In {\it \bibinfo{booktitle}{Proceedings of the 2014 ACM International
  Joint Conference on Pervasive and Ubiquitous Computing: Adjunct
  Publication}\/} (pp. \bibinfo{pages}{589--593}).
\bibitem[{Kirk et~al.(2021)Kirk, Catal \& Tekinerdogan}]{KIRK2021104365}
\bibinfo{author}{Kirk, D.}, \bibinfo{author}{Catal, C.}, \&
  \bibinfo{author}{Tekinerdogan, B.} (\bibinfo{year}{2021}).
\newblock \bibinfo{title}{Precision nutrition: A systematic literature review}.
\newblock {\it \bibinfo{journal}{Computers in Biology and Medicine}\/},  {\it
  \bibinfo{volume}{133}\/}, \bibinfo{pages}{104365}.
  \DOIprefix\doi{10.1016/j.compbiomed.2021.104365}.
\bibitem[{Knez \& Šajn(2020)}]{KNEZ2020460}
\bibinfo{author}{Knez, S.}, \& \bibinfo{author}{Šajn, L.}
  (\bibinfo{year}{2020}).
\newblock \bibinfo{title}{Food object recognition using a mobile device:
  Evaluation of currently implemented systems}.
\newblock {\it \bibinfo{journal}{Trends in Food Science \& Technology}\/},
  {\it \bibinfo{volume}{99}\/}, \bibinfo{pages}{460--471}.
  \DOIprefix\doi{10.1016/j.tifs.2020.03.017}.
\bibitem[{Koenigstorfer et~al.(2014)Koenigstorfer, Groeppel-Klein \&
  Kamm}]{koenigstorfer2014healthful}
\bibinfo{author}{Koenigstorfer, J.}, \bibinfo{author}{Groeppel-Klein, A.}, \&
  \bibinfo{author}{Kamm, F.} (\bibinfo{year}{2014}).
\newblock \bibinfo{title}{Healthful food decision making in response to traffic
  light color-coded nutrition labeling}.
\newblock {\it \bibinfo{journal}{Journal of Public Policy \& Marketing}\/},
  {\it \bibinfo{volume}{33}\/}, \bibinfo{pages}{65--77}.
\bibitem[{Koepp et~al.(2020)Koepp, Baron, Martins, Brandenburg, Kira, Trindade,
  Dominguez, Carneiro, Frozza, Possuelo et~al.}]{koepp2020quality}
\bibinfo{author}{Koepp, J.}, \bibinfo{author}{Baron, M.~V.},
  \bibinfo{author}{Martins, P. R.~H.}, \bibinfo{author}{Brandenburg, C.},
  \bibinfo{author}{Kira, A. T.~F.}, \bibinfo{author}{Trindade, V.~D.},
  \bibinfo{author}{Dominguez, L. M.~L.}, \bibinfo{author}{Carneiro, M.},
  \bibinfo{author}{Frozza, R.}, \bibinfo{author}{Possuelo, L.~G.} et~al.
  (\bibinfo{year}{2020}).
\newblock \bibinfo{title}{The quality of mobile apps used for the
  identification of pressure ulcers in adults: Systematic survey and review of
  apps in app stores}.
\newblock {\it \bibinfo{journal}{JMIR mHealth and uHealth}\/},  {\it
  \bibinfo{volume}{8}\/}, \bibinfo{pages}{e14266}.
\bibitem[{Kong \& Tan(2012)}]{kong2012dietcam}
\bibinfo{author}{Kong, F.}, \& \bibinfo{author}{Tan, J.}
  (\bibinfo{year}{2012}).
\newblock \bibinfo{title}{Dietcam: Automatic dietary assessment with mobile
  camera phones}.
\newblock {\it \bibinfo{journal}{Pervasive and Mobile Computing}\/},  {\it
  \bibinfo{volume}{8}\/}, \bibinfo{pages}{147--163}.
\bibitem[{Koo \& Li(2016)}]{koo2016guideline}
\bibinfo{author}{Koo, T.~K.}, \& \bibinfo{author}{Li, M.~Y.}
  (\bibinfo{year}{2016}).
\newblock \bibinfo{title}{A guideline of selecting and reporting intraclass
  correlation coefficients for reliability research}.
\newblock {\it \bibinfo{journal}{Journal of chiropractic medicine}\/},  {\it
  \bibinfo{volume}{15}\/}, \bibinfo{pages}{155--163}.
\bibitem[{Lange(2011)}]{lange2011inter}
\bibinfo{author}{Lange, R.} (\bibinfo{year}{2011}).
\newblock \bibinfo{title}{Inter-rater reliability}.
\newblock {\it \bibinfo{journal}{Encyclopedia of Clinical Neuropsychology. New
  York, NY: Springer New York}\/},  (p. \bibinfo{pages}{1348}).
\bibitem[{Liang et~al.(2019)Liang, Abbott, Howard, Lim, Ward \&
  Elgendi}]{liang2019effective}
\bibinfo{author}{Liang, Y.}, \bibinfo{author}{Abbott, D.},
  \bibinfo{author}{Howard, N.}, \bibinfo{author}{Lim, K.},
  \bibinfo{author}{Ward, R.}, \& \bibinfo{author}{Elgendi, M.}
  (\bibinfo{year}{2019}).
\newblock \bibinfo{title}{How effective is pulse arrival time for evaluating
  blood pressure? challenges and recommendations from a study using the mimic
  database}.
\newblock {\it \bibinfo{journal}{Journal of clinical medicine}\/},  {\it
  \bibinfo{volume}{8}\/}, \bibinfo{pages}{337}.
\bibitem[{Lieffers et~al.(2014)Lieffers, Vance \& Hanning}]{lieffers2014use}
\bibinfo{author}{Lieffers, J.~R.}, \bibinfo{author}{Vance, V.~A.}, \&
  \bibinfo{author}{Hanning, R.~M.} (\bibinfo{year}{2014}).
\newblock \bibinfo{title}{Use of mobile device applications in canadian
  dietetic practice}.
\newblock {\it \bibinfo{journal}{Canadian journal of dietetic practice and
  research}\/},  {\it \bibinfo{volume}{75}\/}, \bibinfo{pages}{41--47}.
\bibitem[{Liu et~al.(2021)Liu, Pu \& Sun}]{LIU2021193}
\bibinfo{author}{Liu, Y.}, \bibinfo{author}{Pu, H.}, \& \bibinfo{author}{Sun,
  D.-W.} (\bibinfo{year}{2021}).
\newblock \bibinfo{title}{Efficient extraction of deep image features using
  convolutional neural network (cnn) for applications in detecting and
  analysing complex food matrices}.
\newblock {\it \bibinfo{journal}{Trends in Food Science \& Technology}\/},
  {\it \bibinfo{volume}{113}\/}, \bibinfo{pages}{193--204}.
  \DOIprefix\doi{10.1016/j.tifs.2021.04.042}.
\bibitem[{Mai \& Hoffmann(2017)}]{mai2017indirect}
\bibinfo{author}{Mai, R.}, \& \bibinfo{author}{Hoffmann, S.}
  (\bibinfo{year}{2017}).
\newblock \bibinfo{title}{Indirect ways to foster healthier food consumption
  patterns: Health-supportive side effects of health-unrelated motives}.
\newblock {\it \bibinfo{journal}{Food Quality and Preference}\/},  {\it
  \bibinfo{volume}{57}\/}, \bibinfo{pages}{54--68}.
\bibitem[{Marin et~al.(2019)Marin, Biswas, Ofli, Hynes, Salvador, Aytar, Weber
  \& Torralba}]{marin2019recipe1m+}
\bibinfo{author}{Marin, J.}, \bibinfo{author}{Biswas, A.},
  \bibinfo{author}{Ofli, F.}, \bibinfo{author}{Hynes, N.},
  \bibinfo{author}{Salvador, A.}, \bibinfo{author}{Aytar, Y.},
  \bibinfo{author}{Weber, I.}, \& \bibinfo{author}{Torralba, A.}
  (\bibinfo{year}{2019}).
\newblock \bibinfo{title}{Recipe1m+: A dataset for learning cross-modal
  embeddings for cooking recipes and food images}.
\newblock {\it \bibinfo{journal}{IEEE transactions on pattern analysis and
  machine intelligence}\/},  {\it \bibinfo{volume}{43}\/},
  \bibinfo{pages}{187--203}.
\bibitem[{McAllister et~al.(2018)McAllister, Zheng, Bond \&
  Moorhead}]{mcallister2018combining}
\bibinfo{author}{McAllister, P.}, \bibinfo{author}{Zheng, H.},
  \bibinfo{author}{Bond, R.}, \& \bibinfo{author}{Moorhead, A.}
  (\bibinfo{year}{2018}).
\newblock \bibinfo{title}{Combining deep residual neural network features with
  supervised machine learning algorithms to classify diverse food image
  datasets}.
\newblock {\it \bibinfo{journal}{Computers in biology and medicine}\/},  {\it
  \bibinfo{volume}{95}\/}, \bibinfo{pages}{217--233}.
\bibitem[{Meyers et~al.(2015)Meyers, Johnston, Rathod, Korattikara, Gorban,
  Silberman, Guadarrama, Papandreou, Huang \& Murphy}]{meyers2015im2calories}
\bibinfo{author}{Meyers, A.}, \bibinfo{author}{Johnston, N.},
  \bibinfo{author}{Rathod, V.}, \bibinfo{author}{Korattikara, A.},
  \bibinfo{author}{Gorban, A.}, \bibinfo{author}{Silberman, N.},
  \bibinfo{author}{Guadarrama, S.}, \bibinfo{author}{Papandreou, G.},
  \bibinfo{author}{Huang, J.}, \& \bibinfo{author}{Murphy, K.~P.}
  (\bibinfo{year}{2015}).
\newblock \bibinfo{title}{Im2calories: towards an automated mobile vision food
  diary}.
\newblock In {\it \bibinfo{booktitle}{Proceedings of the IEEE International
  Conference on Computer Vision}\/} (pp. \bibinfo{pages}{1233--1241}).
\bibitem[{Mezgec \& Korou{\v{s}}i{\'c}~Seljak(2017)}]{mezgec2017nutrinet}
\bibinfo{author}{Mezgec, S.}, \& \bibinfo{author}{Korou{\v{s}}i{\'c}~Seljak,
  B.} (\bibinfo{year}{2017}).
\newblock \bibinfo{title}{Nutrinet: a deep learning food and drink image
  recognition system for dietary assessment}.
\newblock {\it \bibinfo{journal}{Nutrients}\/},  {\it \bibinfo{volume}{9}\/},
  \bibinfo{pages}{657}. \DOIprefix\doi{10.3390/nu9070657}.
\bibitem[{Mezgec \& Seljak(2021)}]{mezgec2021deep}
\bibinfo{author}{Mezgec, S.}, \& \bibinfo{author}{Seljak, B.~K.}
  (\bibinfo{year}{2021}).
\newblock \bibinfo{title}{Deep neural networks for image-based dietary
  assessment}.
\newblock {\it \bibinfo{journal}{Journal of visualized experiments : JoVE}\/},
  . \DOIprefix\doi{10.3791/61906}.
\bibitem[{Michel \& Burbidge(2019)}]{MICHEL2019194}
\bibinfo{author}{Michel, M.}, \& \bibinfo{author}{Burbidge, A.}
  (\bibinfo{year}{2019}).
\newblock \bibinfo{title}{Nutrition in the digital age - how digital tools can
  help to solve the personalized nutrition conundrum}.
\newblock {\it \bibinfo{journal}{Trends in Food Science \& Technology}\/},
  {\it \bibinfo{volume}{90}\/}, \bibinfo{pages}{194--200}.
  \DOIprefix\doi{10.1016/j.tifs.2019.02.018}.
\bibitem[{Millen et~al.(2014)Millen, Wolongevicz, Nonas \&
  Lichtenstein}]{millen20142013}
\bibinfo{author}{Millen, B.~E.}, \bibinfo{author}{Wolongevicz, D.~M.},
  \bibinfo{author}{Nonas, C.~A.}, \& \bibinfo{author}{Lichtenstein, A.~H.}
  (\bibinfo{year}{2014}).
\newblock \bibinfo{title}{2013 american heart association/american college of
  cardiology/the obesity society guideline for the management of overweight and
  obesity in adults: implications and new opportunities for registered
  dietitian nutritionists}.
\newblock {\it \bibinfo{journal}{Journal of the Academy of Nutrition and
  Dietetics}\/},  {\it \bibinfo{volume}{114}\/}, \bibinfo{pages}{1730--1735}.
\bibitem[{Min et~al.(2020)Min, Jiang \& Jain}]{8930090}
\bibinfo{author}{Min, W.}, \bibinfo{author}{Jiang, S.}, \&
  \bibinfo{author}{Jain, R.} (\bibinfo{year}{2020}).
\newblock \bibinfo{title}{Food recommendation: Framework, existing solutions,
  and challenges}.
\newblock {\it \bibinfo{journal}{IEEE Transactions on Multimedia}\/},  {\it
  \bibinfo{volume}{22}\/}, \bibinfo{pages}{2659--2671}.
  \DOIprefix\doi{10.1109/TMM.2019.2958761}.
\bibitem[{Min et~al.(2019)Min, Jiang, Liu, Rui \& Jain}]{min2019survey}
\bibinfo{author}{Min, W.}, \bibinfo{author}{Jiang, S.}, \bibinfo{author}{Liu,
  L.}, \bibinfo{author}{Rui, Y.}, \& \bibinfo{author}{Jain, R.}
  (\bibinfo{year}{2019}).
\newblock \bibinfo{title}{A survey on food computing}.
\newblock {\it \bibinfo{journal}{ACM Computing Surveys (CSUR)}\/},  {\it
  \bibinfo{volume}{52}\/}, \bibinfo{pages}{1--36}.
  \DOIprefix\doi{10.1145/3329168}.
\bibitem[{Ming et~al.(2018)Ming, Chen, Cao, Forde, Ngo \& Chua}]{ming2018food}
\bibinfo{author}{Ming, Z.-Y.}, \bibinfo{author}{Chen, J.},
  \bibinfo{author}{Cao, Y.}, \bibinfo{author}{Forde, C.}, \bibinfo{author}{Ngo,
  C.-W.}, \& \bibinfo{author}{Chua, T.~S.} (\bibinfo{year}{2018}).
\newblock \bibinfo{title}{Food photo recognition for dietary tracking: System
  and experiment}.
\newblock In {\it \bibinfo{booktitle}{International Conference on Multimedia
  Modeling}\/} (pp. \bibinfo{pages}{129--141}).
\newblock \bibinfo{organization}{Springer}.
\bibitem[{Mokdara et~al.(2018)Mokdara, Pusawiro \&
  Harnsomburana}]{mokdara2018personalized}
\bibinfo{author}{Mokdara, T.}, \bibinfo{author}{Pusawiro, P.}, \&
  \bibinfo{author}{Harnsomburana, J.} (\bibinfo{year}{2018}).
\newblock \bibinfo{title}{Personalized food recommendation using deep neural
  network}.
\newblock In {\it \bibinfo{booktitle}{2018 Seventh ICT International Student
  Project Conference (ICT-ISPC)}\/} (pp. \bibinfo{pages}{1--4}).
\newblock \bibinfo{organization}{IEEE}.
\bibitem[{Nag et~al.(2017)Nag, Pandey, Sharma, Lam, Wang \&
  Jain}]{nag2017pocket}
\bibinfo{author}{Nag, N.}, \bibinfo{author}{Pandey, V.},
  \bibinfo{author}{Sharma, A.}, \bibinfo{author}{Lam, J.},
  \bibinfo{author}{Wang, R.}, \& \bibinfo{author}{Jain, R.}
  (\bibinfo{year}{2017}).
\newblock \bibinfo{title}{Pocket dietitian: Automated healthy dish
  recommendations by location}.
\newblock In {\it \bibinfo{booktitle}{International conference on image
  analysis and processing}\/} (pp. \bibinfo{pages}{444--452}).
\newblock \bibinfo{organization}{Springer}.
\bibitem[{Nayak et~al.(2020)Nayak, Vakula, Dinesh, Naik \&
  Pelusi}]{NAYAK2020100297}
\bibinfo{author}{Nayak, J.}, \bibinfo{author}{Vakula, K.},
  \bibinfo{author}{Dinesh, P.}, \bibinfo{author}{Naik, B.}, \&
  \bibinfo{author}{Pelusi, D.} (\bibinfo{year}{2020}).
\newblock \bibinfo{title}{Intelligent food processing: Journey from artificial
  neural network to deep learning}.
\newblock {\it \bibinfo{journal}{Computer Science Review}\/},  {\it
  \bibinfo{volume}{38}\/}, \bibinfo{pages}{100297}.
  \DOIprefix\doi{10.1016/j.cosrev.2020.100297}.
\bibitem[{Ng et~al.(2014)Ng, Fleming, Robinson, Thomson, Graetz, Margono,
  Mullany, Biryukov, Abbafati, Abera et~al.}]{ng2014global}
\bibinfo{author}{Ng, M.}, \bibinfo{author}{Fleming, T.},
  \bibinfo{author}{Robinson, M.}, \bibinfo{author}{Thomson, B.},
  \bibinfo{author}{Graetz, N.}, \bibinfo{author}{Margono, C.},
  \bibinfo{author}{Mullany, E.~C.}, \bibinfo{author}{Biryukov, S.},
  \bibinfo{author}{Abbafati, C.}, \bibinfo{author}{Abera, S.~F.} et~al.
  (\bibinfo{year}{2014}).
\newblock \bibinfo{title}{Global, regional, and national prevalence of
  overweight and obesity in children and adults during 1980--2013: a systematic
  analysis for the global burden of disease study 2013}.
\newblock {\it \bibinfo{journal}{The lancet}\/},  {\it
  \bibinfo{volume}{384}\/}, \bibinfo{pages}{766--781}.
\bibitem[{Pandey et~al.(2017)Pandey, Deepthi, Mandal \&
  Puhan}]{pandey2017foodnet}
\bibinfo{author}{Pandey, P.}, \bibinfo{author}{Deepthi, A.},
  \bibinfo{author}{Mandal, B.}, \& \bibinfo{author}{Puhan, N.~B.}
  (\bibinfo{year}{2017}).
\newblock \bibinfo{title}{Foodnet: Recognizing foods using ensemble of deep
  networks}.
\newblock {\it \bibinfo{journal}{IEEE Signal Processing Letters}\/},  {\it
  \bibinfo{volume}{24}\/}, \bibinfo{pages}{1758--1762}.
\bibitem[{Phanich et~al.(2010)Phanich, Pholkul \&
  Phimoltares}]{phanich2010food}
\bibinfo{author}{Phanich, M.}, \bibinfo{author}{Pholkul, P.}, \&
  \bibinfo{author}{Phimoltares, S.} (\bibinfo{year}{2010}).
\newblock \bibinfo{title}{Food recommendation system using clustering analysis
  for diabetic patients}.
\newblock In {\it \bibinfo{booktitle}{2010 International Conference on
  Information Science and Applications}\/} (pp. \bibinfo{pages}{1--8}).
\newblock \bibinfo{organization}{IEEE}.
\bibitem[{Poon \& Friesen(2015)}]{poon2015algorithms}
\bibinfo{author}{Poon, T. W.~K.}, \& \bibinfo{author}{Friesen, M.~R.}
  (\bibinfo{year}{2015}).
\newblock \bibinfo{title}{Algorithms for size and color detection of smartphone
  images of chronic wounds for healthcare applications}.
\newblock {\it \bibinfo{journal}{IEEE Access}\/},  {\it \bibinfo{volume}{3}\/},
  \bibinfo{pages}{1799--1808}.
\bibitem[{Pouladzadeh et~al.(2014)Pouladzadeh, Shirmohammadi \&
  Al-Maghrabi}]{pouladzadeh2014measuring}
\bibinfo{author}{Pouladzadeh, P.}, \bibinfo{author}{Shirmohammadi, S.}, \&
  \bibinfo{author}{Al-Maghrabi, R.} (\bibinfo{year}{2014}).
\newblock \bibinfo{title}{Measuring calorie and nutrition from food image}.
\newblock {\it \bibinfo{journal}{IEEE Transactions on Instrumentation and
  Measurement}\/},  {\it \bibinfo{volume}{63}\/}, \bibinfo{pages}{1947--1956}.
\bibitem[{Pritha et~al.(2021)Pritha, Tasnim, Kabir, Amin \&
  Das}]{pritha2021systematic}
\bibinfo{author}{Pritha, S.~T.}, \bibinfo{author}{Tasnim, R.},
  \bibinfo{author}{Kabir, M.~A.}, \bibinfo{author}{Amin, S.}, \&
  \bibinfo{author}{Das, A.} (\bibinfo{year}{2021}).
\newblock \bibinfo{title}{A systematic review of mobile apps for child sexual
  abuse education: Limitations and design guidelines}.
\newblock \href{http://arxiv.org/abs/2107.01596}{\tt arXiv:2107.01596}.
\bibitem[{Puri et~al.(2009)Puri, Zhu, Yu, Divakaran \&
  Sawhney}]{puri2009recognition}
\bibinfo{author}{Puri, M.}, \bibinfo{author}{Zhu, Z.}, \bibinfo{author}{Yu,
  Q.}, \bibinfo{author}{Divakaran, A.}, \& \bibinfo{author}{Sawhney, H.}
  (\bibinfo{year}{2009}).
\newblock \bibinfo{title}{Recognition and volume estimation of food intake
  using a mobile device}.
\newblock In {\it \bibinfo{booktitle}{2009 Workshop on Applications of Computer
  Vision (WACV)}\/} (pp. \bibinfo{pages}{1--8}).
\newblock \bibinfo{organization}{IEEE}.
\bibitem[{Rav{\`\i} et~al.(2015)Rav{\`\i}, Lo \& Yang}]{ravi2015real}
\bibinfo{author}{Rav{\`\i}, D.}, \bibinfo{author}{Lo, B.}, \&
  \bibinfo{author}{Yang, G.-Z.} (\bibinfo{year}{2015}).
\newblock \bibinfo{title}{Real-time food intake classification and energy
  expenditure estimation on a mobile device}.
\newblock In {\it \bibinfo{booktitle}{2015 IEEE 12th International Conference
  on Wearable and Implantable Body Sensor Networks (BSN)}\/} (pp.
  \bibinfo{pages}{1--6}).
\newblock \bibinfo{organization}{IEEE}.
\bibitem[{Rivera et~al.(2016)Rivera, McPherson, Hamilton, Birken, Coons, Iyer,
  Agarwal, Lalloo \& Stinson}]{rivera2016mobile}
\bibinfo{author}{Rivera, J.}, \bibinfo{author}{McPherson, A.},
  \bibinfo{author}{Hamilton, J.}, \bibinfo{author}{Birken, C.},
  \bibinfo{author}{Coons, M.}, \bibinfo{author}{Iyer, S.},
  \bibinfo{author}{Agarwal, A.}, \bibinfo{author}{Lalloo, C.}, \&
  \bibinfo{author}{Stinson, J.} (\bibinfo{year}{2016}).
\newblock \bibinfo{title}{Mobile apps for weight management: a scoping review}.
\newblock {\it \bibinfo{journal}{JMIR mHealth and uHealth}\/},  {\it
  \bibinfo{volume}{4}\/}, \bibinfo{pages}{e87}.
\bibitem[{Rokicki et~al.(2018)Rokicki, Trattner \& Herder}]{rokicki2018impact}
\bibinfo{author}{Rokicki, M.}, \bibinfo{author}{Trattner, C.}, \&
  \bibinfo{author}{Herder, E.} (\bibinfo{year}{2018}).
\newblock \bibinfo{title}{The impact of recipe features, social cues and
  demographics on estimating the healthiness of online recipes}.
\newblock In {\it \bibinfo{booktitle}{Proceedings of the International AAAI
  Conference on Web and Social Media}\/}.
\newblock volume~\bibinfo{volume}{12}.
\bibitem[{Ross \& Gao(2016)}]{ross2016overcoming}
\bibinfo{author}{Ross, J.}, \& \bibinfo{author}{Gao, J.}
  (\bibinfo{year}{2016}).
\newblock \bibinfo{title}{Overcoming the language barrier in mobile user
  interface design: A case study on a mobile health app}.
\newblock {\it \bibinfo{journal}{arXiv preprint arXiv:1605.04693}\/}, .
\bibitem[{Sawa \& Morikawa(2007)}]{sawa2007interrater}
\bibinfo{author}{Sawa, J.}, \& \bibinfo{author}{Morikawa, T.}
  (\bibinfo{year}{2007}).
\newblock \bibinfo{title}{Interrater reliability for multiple raters in
  clinical trials of ordinal scale}.
\newblock {\it \bibinfo{journal}{Drug information journal: DIJ/Drug Information
  Association}\/},  {\it \bibinfo{volume}{41}\/}, \bibinfo{pages}{595--605}.
\bibitem[{Speiser et~al.(2005)Speiser, Rudolf, Anhalt, Camacho-Hubner,
  Chiarelli, Eliakim, Freemark, Gruters, Hershkovitz, Iughetti
  et~al.}]{speiser2005obesity}
\bibinfo{author}{Speiser, P.~W.}, \bibinfo{author}{Rudolf, M.},
  \bibinfo{author}{Anhalt, H.}, \bibinfo{author}{Camacho-Hubner, C.},
  \bibinfo{author}{Chiarelli, F.}, \bibinfo{author}{Eliakim, A.},
  \bibinfo{author}{Freemark, M.}, \bibinfo{author}{Gruters, A.},
  \bibinfo{author}{Hershkovitz, E.}, \bibinfo{author}{Iughetti, L.} et~al.
  (\bibinfo{year}{2005}).
\newblock \bibinfo{title}{Obesity consensus working group childhood obesity}.
\newblock {\it \bibinfo{journal}{J Clin Endocrinol Metab}\/},  {\it
  \bibinfo{volume}{90}\/}, \bibinfo{pages}{1871--87}.
\bibitem[{Sriram et~al.(1996)Sriram, Rao, Biswas \&
  Ahmed}]{sriram1996applications}
\bibinfo{author}{Sriram, T.}, \bibinfo{author}{Rao, K.~V.},
  \bibinfo{author}{Biswas, S.}, \& \bibinfo{author}{Ahmed, B.}
  (\bibinfo{year}{1996}).
\newblock \bibinfo{title}{Applications of barcode technology in automated
  storage and retrieval systems}.
\newblock In {\it \bibinfo{booktitle}{Proceedings of the 1996 IEEE IECON. 22nd
  International Conference on Industrial Electronics, Control, and
  Instrumentation}\/} (pp. \bibinfo{pages}{641--646}).
\newblock \bibinfo{organization}{IEEE} volume~\bibinfo{volume}{1}.
\bibitem[{Stawarz et~al.(2015)Stawarz, Cox \& Blandford}]{stawarz2015beyond}
\bibinfo{author}{Stawarz, K.}, \bibinfo{author}{Cox, A.~L.}, \&
  \bibinfo{author}{Blandford, A.} (\bibinfo{year}{2015}).
\newblock \bibinfo{title}{Beyond self-tracking and reminders: designing
  smartphone apps that support habit formation}.
\newblock In {\it \bibinfo{booktitle}{Proceedings of the 33rd annual ACM
  conference on human factors in computing systems}\/} (pp.
  \bibinfo{pages}{2653--2662}).
\newblock \bibinfo{publisher}{ACM}.
\bibitem[{Stoyanov et~al.(2016)Stoyanov, Hides, Kavanagh \&
  Wilson}]{stoyanov2016mobile}
\bibinfo{author}{Stoyanov, S.}, \bibinfo{author}{Hides, L.},
  \bibinfo{author}{Kavanagh, D.}, \& \bibinfo{author}{Wilson, H.}
  (\bibinfo{year}{2016}).
\newblock \bibinfo{title}{Development and validation of the user version of the
  mobile application rating scale (umars)}.
\newblock {\it \bibinfo{journal}{JMIR Mhealth Uhealth}\/},  {\it
  \bibinfo{volume}{4}\/}, \bibinfo{pages}{e72}.
  \DOIprefix\doi{10.2196/mhealth.5849}.
\bibitem[{Stoyanov et~al.(2015)Stoyanov, Hides, Kavanagh, Zelenko,
  Tjondronegoro \& Mani}]{stoyanov2015mobile}
\bibinfo{author}{Stoyanov, S.~R.}, \bibinfo{author}{Hides, L.},
  \bibinfo{author}{Kavanagh, D.~J.}, \bibinfo{author}{Zelenko, O.},
  \bibinfo{author}{Tjondronegoro, D.}, \& \bibinfo{author}{Mani, M.}
  (\bibinfo{year}{2015}).
\newblock \bibinfo{title}{Mobile app rating scale: a new tool for assessing the
  quality of health mobile apps}.
\newblock {\it \bibinfo{journal}{JMIR mHealth and uHealth}\/},  {\it
  \bibinfo{volume}{3}\/}, \bibinfo{pages}{e27}.
\bibitem[{Sun et~al.(2010)Sun, Fernstrom, Jia, Hackworth, Yao, Li, Li,
  Fernstrom \& Sclabassi}]{sun2010wearable}
\bibinfo{author}{Sun, M.}, \bibinfo{author}{Fernstrom, J.~D.},
  \bibinfo{author}{Jia, W.}, \bibinfo{author}{Hackworth, S.~A.},
  \bibinfo{author}{Yao, N.}, \bibinfo{author}{Li, Y.}, \bibinfo{author}{Li,
  C.}, \bibinfo{author}{Fernstrom, M.~H.}, \& \bibinfo{author}{Sclabassi,
  R.~J.} (\bibinfo{year}{2010}).
\newblock \bibinfo{title}{A wearable electronic system for objective dietary
  assessment}.
\newblock {\it \bibinfo{journal}{Journal of the American Dietetic
  Association}\/},  {\it \bibinfo{volume}{110}\/}, \bibinfo{pages}{45--47}.
\bibitem[{Tahir \& Loo(2021)}]{healthcare9121676}
\bibinfo{author}{Tahir, G.~A.}, \& \bibinfo{author}{Loo, C.~K.}
  (\bibinfo{year}{2021}).
\newblock \bibinfo{title}{A comprehensive survey of image-based food
  recognition and volume estimation methods for dietary assessment}.
\newblock {\it \bibinfo{journal}{Healthcare}\/},  {\it \bibinfo{volume}{9}\/}.
  \DOIprefix\doi{10.3390/healthcare9121676}.
\bibitem[{Tanno et~al.(2016)Tanno, Okamoto \& Yanai}]{tanno2016deepfoodcam}
\bibinfo{author}{Tanno, R.}, \bibinfo{author}{Okamoto, K.}, \&
  \bibinfo{author}{Yanai, K.} (\bibinfo{year}{2016}).
\newblock \bibinfo{title}{Deepfoodcam: A dcnn-based real-time mobile food
  recognition system}.
\newblock In {\it \bibinfo{booktitle}{Proceedings of the 2nd International
  Workshop on Multimedia Assisted Dietary Management}\/} (pp.
  \bibinfo{pages}{89--89}).
\bibitem[{Tay et~al.(2020)Tay, Kaur, Quek, Lim \& Henry}]{tay2020current}
\bibinfo{author}{Tay, W.}, \bibinfo{author}{Kaur, B.}, \bibinfo{author}{Quek,
  R.}, \bibinfo{author}{Lim, J.}, \& \bibinfo{author}{Henry, C.~J.}
  (\bibinfo{year}{2020}).
\newblock \bibinfo{title}{Current developments in digital quantitative volume
  estimation for the optimisation of dietary assessment}.
\newblock {\it \bibinfo{journal}{Nutrients}\/},  {\it \bibinfo{volume}{12}\/},
  \bibinfo{pages}{1167}.
\bibitem[{Tricco et~al.(2018)Tricco, Lillie, Zarin, O'Brien, Colquhoun, Levac,
  Moher, Peters, Horsley, Weeks et~al.}]{tricco2018prisma}
\bibinfo{author}{Tricco, A.~C.}, \bibinfo{author}{Lillie, E.},
  \bibinfo{author}{Zarin, W.}, \bibinfo{author}{O'Brien, K.~K.},
  \bibinfo{author}{Colquhoun, H.}, \bibinfo{author}{Levac, D.},
  \bibinfo{author}{Moher, D.}, \bibinfo{author}{Peters, M.~D.},
  \bibinfo{author}{Horsley, T.}, \bibinfo{author}{Weeks, L.} et~al.
  (\bibinfo{year}{2018}).
\newblock \bibinfo{title}{Prisma extension for scoping reviews (prisma-scr):
  checklist and explanation}.
\newblock {\it \bibinfo{journal}{Annals of internal medicine}\/},  {\it
  \bibinfo{volume}{169}\/}, \bibinfo{pages}{467--473}.
\bibitem[{Ursachi et~al.(2015)Ursachi, Horodnic \& Zait}]{ursachi2015reliable}
\bibinfo{author}{Ursachi, G.}, \bibinfo{author}{Horodnic, I.~A.}, \&
  \bibinfo{author}{Zait, A.} (\bibinfo{year}{2015}).
\newblock \bibinfo{title}{How reliable are measurement scales? external factors
  with indirect influence on reliability estimators}.
\newblock {\it \bibinfo{journal}{Procedia Economics and Finance}\/},  {\it
  \bibinfo{volume}{20}\/}, \bibinfo{pages}{679--686}.
\bibitem[{Van~Asbroeck \& Matthys(2020)}]{van2020use}
\bibinfo{author}{Van~Asbroeck, S.}, \& \bibinfo{author}{Matthys, C.}
  (\bibinfo{year}{2020}).
\newblock \bibinfo{title}{Use of different food image recognition platforms in
  dietary assessment: Comparison study}.
\newblock {\it \bibinfo{journal}{JMIR formative research}\/},  {\it
  \bibinfo{volume}{4}\/}, \bibinfo{pages}{e15602}.
  \DOIprefix\doi{10.2196/15602}.
\bibitem[{Vasa et~al.(2012)Vasa, Hoon, Mouzakis \&
  Noguchi}]{vasa2012preliminary}
\bibinfo{author}{Vasa, R.}, \bibinfo{author}{Hoon, L.},
  \bibinfo{author}{Mouzakis, K.}, \& \bibinfo{author}{Noguchi, A.}
  (\bibinfo{year}{2012}).
\newblock \bibinfo{title}{A preliminary analysis of mobile app user reviews}.
\newblock In {\it \bibinfo{booktitle}{Proceedings of the 24th Australian
  computer-human interaction conference}\/} (pp. \bibinfo{pages}{241--244}).
\bibitem[{Vivek et~al.(2018)Vivek, Manju \& Vijay}]{vivek2018machine}
\bibinfo{author}{Vivek, M.}, \bibinfo{author}{Manju, N.}, \&
  \bibinfo{author}{Vijay, M.} (\bibinfo{year}{2018}).
\newblock \bibinfo{title}{Machine learning based food recipe recommendation
  system}.
\newblock In {\it \bibinfo{booktitle}{Proceedings of international conference
  on cognition and recognition}\/} (pp. \bibinfo{pages}{11--19}).
\newblock \bibinfo{organization}{Springer}.
\bibitem[{Vos-Draper(2013)}]{vos2013poster}
\bibinfo{author}{Vos-Draper, T.} (\bibinfo{year}{2013}).
\newblock \bibinfo{title}{Poster 29 wireless seat interface pressure mapping on
  a smartphone: feasibility study in users with sci}.
\newblock {\it \bibinfo{journal}{Archives of Physical Medicine and
  Rehabilitation}\/},  {\it \bibinfo{volume}{94}\/}, \bibinfo{pages}{e21--e22}.
\bibitem[{Wang et~al.(2021)Wang, Duan, Jiang, Jing, Song \&
  Nie}]{10.1145/3418211}
\bibinfo{author}{Wang, W.}, \bibinfo{author}{Duan, L.-Y.},
  \bibinfo{author}{Jiang, H.}, \bibinfo{author}{Jing, P.},
  \bibinfo{author}{Song, X.}, \& \bibinfo{author}{Nie, L.}
  (\bibinfo{year}{2021}).
\newblock \bibinfo{title}{Market2dish: Health-aware food recommendation}.
\newblock {\it \bibinfo{journal}{ACM Trans. Multimedia Comput. Commun.
  Appl.}\/},  {\it \bibinfo{volume}{17}\/}. \DOIprefix\doi{10.1145/3418211}.
\bibitem[{Wu(2007)}]{wu2007empirical}
\bibinfo{author}{Wu, C.-H.} (\bibinfo{year}{2007}).
\newblock \bibinfo{title}{An empirical study on the transformation of
  likert-scale data to numerical scores}.
\newblock {\it \bibinfo{journal}{Applied Mathematical Sciences}\/},  {\it
  \bibinfo{volume}{1}\/}, \bibinfo{pages}{2851--2862}.
\bibitem[{Wu et~al.(2016)Wu, Merler, Uceda-Sosa \& Smith}]{wu2016learning}
\bibinfo{author}{Wu, H.}, \bibinfo{author}{Merler, M.},
  \bibinfo{author}{Uceda-Sosa, R.}, \& \bibinfo{author}{Smith, J.~R.}
  (\bibinfo{year}{2016}).
\newblock \bibinfo{title}{Learning to make better mistakes: Semantics-aware
  visual food recognition}.
\newblock In {\it \bibinfo{booktitle}{Proceedings of the 24th ACM international
  conference on Multimedia}\/} (pp. \bibinfo{pages}{172--176}).
\bibitem[{Yang et~al.(2017)Yang, Hsieh, Yang, Pollak, Dell, Belongie, Cole \&
  Estrin}]{yang2017yum}
\bibinfo{author}{Yang, L.}, \bibinfo{author}{Hsieh, C.-K.},
  \bibinfo{author}{Yang, H.}, \bibinfo{author}{Pollak, J.~P.},
  \bibinfo{author}{Dell, N.}, \bibinfo{author}{Belongie, S.},
  \bibinfo{author}{Cole, C.}, \& \bibinfo{author}{Estrin, D.}
  (\bibinfo{year}{2017}).
\newblock \bibinfo{title}{Yum-me: a personalized nutrient-based meal
  recommender system}.
\newblock {\it \bibinfo{journal}{ACM Transactions on Information Systems
  (TOIS)}\/},  {\it \bibinfo{volume}{36}\/}, \bibinfo{pages}{1--31}.
\bibitem[{Yang et~al.(2010)Yang, Chen, Pomerleau \& Sukthankar}]{yang2010food}
\bibinfo{author}{Yang, S.}, \bibinfo{author}{Chen, M.},
  \bibinfo{author}{Pomerleau, D.}, \& \bibinfo{author}{Sukthankar, R.}
  (\bibinfo{year}{2010}).
\newblock \bibinfo{title}{Food recognition using statistics of pairwise local
  features}.
\newblock In {\it \bibinfo{booktitle}{2010 IEEE Computer Society Conference on
  Computer Vision and Pattern Recognition}\/} (pp.
  \bibinfo{pages}{2249--2256}).
\newblock \bibinfo{organization}{IEEE}.
\bibitem[{Yang et~al.(2019)Yang, Jia, Bucher, Zhang \& Sun}]{yang2019image}
\bibinfo{author}{Yang, Y.}, \bibinfo{author}{Jia, W.}, \bibinfo{author}{Bucher,
  T.}, \bibinfo{author}{Zhang, H.}, \& \bibinfo{author}{Sun, M.}
  (\bibinfo{year}{2019}).
\newblock \bibinfo{title}{Image-based food portion size estimation using a
  smartphone without a fiducial marker}.
\newblock {\it \bibinfo{journal}{Public health nutrition}\/},  {\it
  \bibinfo{volume}{22}\/}, \bibinfo{pages}{1180--1192}.
\bibitem[{Zhang et~al.(2015)Zhang, Yu, Siddiquie, Divakaran \&
  Sawhney}]{zhang2015snap}
\bibinfo{author}{Zhang, W.}, \bibinfo{author}{Yu, Q.},
  \bibinfo{author}{Siddiquie, B.}, \bibinfo{author}{Divakaran, A.}, \&
  \bibinfo{author}{Sawhney, H.} (\bibinfo{year}{2015}).
\newblock \bibinfo{title}{“snap-n-eat” food recognition and nutrition
  estimation on a smartphone}.
\newblock {\it \bibinfo{journal}{Journal of diabetes science and
  technology}\/},  {\it \bibinfo{volume}{9}\/}, \bibinfo{pages}{525--533}.
\bibitem[{Zheng et~al.(2017)Zheng, Wang \& Zhu}]{zheng2017food}
\bibinfo{author}{Zheng, J.}, \bibinfo{author}{Wang, Z.~J.}, \&
  \bibinfo{author}{Zhu, C.} (\bibinfo{year}{2017}).
\newblock \bibinfo{title}{Food image recognition via superpixel based low-level
  and mid-level distance coding for smart home applications}.
\newblock {\it \bibinfo{journal}{Sustainability}\/},  {\it
  \bibinfo{volume}{9}\/}, \bibinfo{pages}{856}.

\end{thebibliography}






\appendix
\section{Extended food consumption tracking and recommendation app rating domains and criteria}
\label{sec:sample:appendix}
\begin{longtable}{l l}
\caption{Extended food consumption tracking and recommendation app rating domains and criteria.}
\label{domain}\\
\hline
Sub-scale & Criteria\\
\hline
\hline
\endfirsthead
\caption{Extended food consumption tracking and recommendation app rating domains and criteria (continued)}\\
\hline
Domain & Criteria\\ 
\hline
\endhead
\hline 
\endfoot
\hline
\endlastfoot
\hline
\multirow{7}{*}{App metadata} 
&App platform \\
&App store rating \\ 
&App store description \\
&App store URL\\
&Number of downloads \\
& Origin \\
& Developer  \\
\hline

\multirow{3}{*}{App classification} 
&App sub-category \\
&Applicable age groups \\ 
&App price \\

\hline
\multirow{4}{*}{Aesthetics} 
&Layout consistency and readability\\
&Content resolution \\ 
&Visual appeal \\
& Group targeting according to app content\\

\hline
\multirow{7}{*}{General app features} 
&Social sharing feature\\
&Authentication feature\\
&User on boarding interfaces\\
&Content customization\\
&Visual information\\
&Data export options\\
&Subscription options\\

\hline
\multirow{6}{*}{Performance and efficiency}
&Bootup efficiency\\
&Accuracy of features and components\\
&Responsiveness of app\\
&Frequency of app crash\\
&Overheating device issues\\
&Battery life impact\\

\hline
\multirow{4}{*}{Usability}
&Ease of Use\\
&Navigational accuracy\\
&Gestural design\\
&Interactivity and user feedback\\
\hline

\multirow{6}{*}{App specific functionality} 
&Recognition of food\\ 
&Volume computation\\
&Nutrition value estimation\\
&Visualization of food consumption history\\
&Recommendation of food\\
&New addition of food items\\
\hline

\multirow{4}{*}{Subjective quality} 
&Overall app purchase preference\\ 
&Overall app recommendation\\
&Frequency of use based on relevance\\
& Overall star rating\\
\hline

\multirow{5}{*}{Transparency} 
& Accuracy of store description\\
& Credibility/legitimacy of source\\
& Verification by evidence \\
& Feasibility of achieving goals \\
& User consent\\
\hline
\multirow{6}{*}{Perceived impact of app on users} 
&Awareness induction behavior \\
&Knowledge enhancing behavior \\ 
& Change of attitude toward improving balanced diet \\
&Intention to change\\
&Balanced diet related help-seeking behaviour \\
& Behaviour change \\
\hline

\end{longtable}

\end{document}